\documentclass[acmsmall,screen]{acmart}
\acmJournal{TOSEM}

\usepackage{multirow}
\usepackage{tabularx}
\usepackage{tabulary}
\usepackage{rotating}
\usepackage{lscape}
\usepackage{tcolorbox}
\usepackage{xcolor}

\usepackage{caption}
\usepackage{subcaption}
\usepackage{ifthen}
\newboolean{hidecomments}
\setboolean{hidecomments}{false}
\ifthenelse{\boolean{hidecomments}}
{\newcommand{\nb}[2]{}}
{\newcommand{\nb}[2]{
    \fbox{\bfseries\sffamily\scriptsize#1}
    {\sf\small$\blacktriangleright$ 
      {#2} $\blacktriangleleft$}}} 

\newcolumntype{L}[1]{>{\raggedright\let\newline\\\arraybackslash\hspace{0pt}}m{#1}}
\newcolumntype{C}[1]{>{\centering\let\newline\\\arraybackslash\hspace{0pt}}m{#1}}
\newcolumntype{R}[1]{>{\raggedleft\let\newline\\\arraybackslash\hspace{0pt}}m{#1}}
\definecolor{ao(english)}{rgb}{0.0, 0.5, 0.0}
\usepackage{xcolor, soul}
\sethlcolor{lightgray}

\usepackage{color}
\usepackage{listings}
\definecolor{GrayCodeBlock}{RGB}{241,241,241}
\definecolor{BlackText}{RGB}{110,107,94}
\definecolor{RedTypename}{RGB}{182,86,17}
\definecolor{GreenString}{RGB}{96,172,57}
\definecolor{PurpleKeyword}{RGB}{184,84,212}
\definecolor{GrayComment}{RGB}{170,170,170}
\definecolor{GoldDocumentation}{RGB}{180,165,45}
\lstdefinelanguage{rust}
{
    columns=fullflexible,
    keepspaces=true,
    frame=single,
    framesep=0pt,
    framerule=0pt,
    framexleftmargin=4pt,
    framexrightmargin=4pt,
    framextopmargin=5pt,
    framexbottommargin=3pt,
    xleftmargin=4pt,
    xrightmargin=4pt,
    backgroundcolor=\color{GrayCodeBlock},
    basicstyle=\ttfamily\color{BlackText},
    keywords={
        true,false,
        unsafe,async,await,move,
        use,pub,crate,super,self,mod,
        struct,enum,fn,const,static,let,mut,ref,type,impl,dyn,trait,where,as,
        break,continue,if,else,while,for,loop,match,return,yield,in
    },
   keywordstyle=\color{GreenString},
ndkeywords={
        bool,u8,u16,u32,u64,u128,i8,i16,i32,i64,i128,char,str,
        Self,Option,Some,None,Result,Ok,Err,String,Box,Vec,Rc,Arc,Cell,RefCell,HashMap,BTreeMap,
        macro_rules
    },
    ndkeywordstyle=\color{RedTypename},
    comment=[l][\color{GrayComment}\slshape]{//},
    morecomment=[s][\color{GrayComment}\slshape]{/*}{*/},
    morecomment=[l][\color{GoldDocumentation}\slshape]{///},
    morecomment=[s][\color{GoldDocumentation}\slshape]{/*!}{*/},
    morecomment=[l][\color{GoldDocumentation}\slshape]{//!},
    morecomment=[s][\color{RedTypename}]{\#![}{]},
    morecomment=[s][\color{RedTypename}]{\#[}{]},
    stringstyle=\color{GreenString},
    string=[b]"
}

\renewcommand\footnotetextcopyrightpermission[1]{} 
\begin{document}

\title{A Closer Look at the Security Risks in the Rust Ecosystem}

\author{Xiaoye Zheng}
\affiliation{%
  \institution{Zhejiang University}
  \city{Hangzhou}
  \country{China}}
\email{xiaoyez@zju.edu.cn}

\author{Zhiyuan Wan}
\authornote{Corresponding author.}
\affiliation{%
  \institution{Zhejiang University}
  \city{Hangzhou}
  \country{China}
}
\email{wanzhiyuan@zju.edu.cn}

\author{Yun Zhang}
\affiliation{%
 \institution{Hangzhou City University}
 \city{Hangzhou}
 \country{China}}
 \email{yunzhang@zucc.edu.cn}

\author{Rui Chang}
\affiliation{%
  \institution{Zhejiang University}
  \city{Hangzhou}
  \country{China}}
\email{crix1021@zju.edu.cn}

\author{David Lo}
\affiliation{%
  \institution{Singapore Management University}
  \country{Singapore}
  }
 \email{davidlo@smu.edu.sg}

\renewcommand{\shortauthors}{Zheng et al.}

\begin{abstract}
Rust is an emerging programming language designed for the development of systems software. To facilitate the reuse of Rust code, {\tt crates.io}, as a central package registry of the Rust ecosystem, hosts thousands of third-party Rust packages. The openness of {\tt crates.io} enables the growth of the Rust ecosystem but comes with security risks by severe security advisories.
Although Rust guarantees a software program to be safe via programming language features and strict compile-time checking, the \textcolor{ao(english)}{\tt unsafe} keyword in Rust allows developers to bypass compiler safety checks for certain regions of code.
Prior studies empirically investigate the memory safety and concurrency bugs in the Rust ecosystem, as well as the usage of \textcolor{ao(english)}{\tt unsafe} keywords in practice.
Nonetheless, the literature lacks a systematic investigation of the  security risks in the Rust ecosystem.

In this paper, we perform a comprehensive investigation into the security risks present in the Rust ecosystem, asking ``what are the characteristics of the vulnerabilities, what are the characteristics of the vulnerable packages, and how are the vulnerabilities fixed in practice?''. To facilitate the study, we first compile a dataset of 433 vulnerabilities, 300 vulnerable code repositories, and 218 vulnerability fix commits in the Rust ecosystem, spanning over 7 years. With the dataset, we characterize the types, life spans, and evolution of the disclosed vulnerabilities. We then characterize the popularity, categorization, and vulnerability density of the vulnerable Rust packages, as well as their versions and code regions affected by the disclosed vulnerabilities. Finally, we characterize the complexity of vulnerability fixes and localities of corresponding code changes, and inspect how practitioners fix vulnerabilities in Rust packages with various localities.
%

We find that memory safety and concurrency issues account for nearly two thirds of the vulnerabilities in the Rust ecosystem. It takes over 2 years for the vulnerabilities to become publicly disclosed, and one third of the vulnerabilities have no fixes committed before their disclosure. In terms of vulnerability density, we observe a continuous upward trend at the package level over time, but a decreasing trend at the code level since August 2020.
In the vulnerable Rust packages, the vulnerable code tends to be localized at the file level, and contains statistically significantly more unsafe functions and blocks than the rest of the code. 
More popular packages tend to have more vulnerabilities, while the less popular packages suffer from vulnerabilities for
more versions.
The vulnerability fix commits tend to be localized to a limited number of lines of code.
Developers tend to address vulnerable safe functions by adding safe functions or lines to them, vulnerable unsafe blocks by removing them, and vulnerable unsafe functions by modifying unsafe trait implementations.
Based on our findings, we discuss implications, provide recommendations for software practitioners, and outline directions for future research.
\end{abstract}


\maketitle

\section{Introduction}\label{sec:introduction}
Modern software systems benefit from reusing code from open source projects, leading to the formation of complex interdependency networks, i.e., software ecosystems~\cite{valiev2018ecosystem}. The reusable code usually takes the form of packages delivered by package management systems, such as npm for JavaScript packages, PyPI for Python packages, and Maven for Java packages. 
In recent years, researchers conduct substantial studies to investigate a variety of aspects of software ecosystems, including their evolution~\cite{decan2019empirical,CSMR2013Revol}, dependencies of packages~\cite{SANER2017dependency,ECSAW2016Dependency,SANER2016dependency} and security risks~\cite{zimmermann2019small,alfadel2021empirical,CCS2017Patches}. 
A few studies make comparisons across software ecosystems, such as the structure~\cite{kikas2017structure} and evolution~\cite{decan2019empirical} of dependencies across software ecosystems.     

Rust is an emerging programming language designed for the development of systems software~\cite{pldi2020rust-8,pldi2020rust-69,pldi2020rust-74}. Over the past few years, Rust has experienced explosive growth and gained popularity~\cite{pldi2020rust-46,pldi2020rust-47,pldi2020rust-48}, especially in developing systems software like operating systems and browsers~\cite{pldi2020rust-55, pldi2020rust-59, pldi2020rust-68, pldi2020rust-71, pldi2020rust-77}. According to the annual developer survey of Stack Overflow\footnote{\url{https://insights.stackoverflow.com/survey/2021\#technology-most-loved-dreaded-and-wanted}}, Rust has been named the ``most loved programming language'' for six years in a row, from  2016 to 2021. 
To support Rust practitioners with third-party code, {\tt crates.io}, as a central package registry of the Rust ecosystem, provides thousands of reusable packages (crates). 
The openness of {\tt crates.io} enables the growth of the Rust ecosystem, ranging from small utility packages to complex Web programming frameworks and cryptography libraries.
Rust guarantees a software program to be safe via programming language features, and with strict compile-time checking~\cite{icse2020evan-15,icse2020evan-29,icse2020evan-30}.
Nonetheless, the openness of the Rust ecosystem comes with security risks as evidenced by severe security advisories. For instance, in January 2022, Rust maintainers released a security update for a high-severity vulnerability (CVE-2022-21658). Attackers could abuse the vulnerability to purge files and directories from a vulnerable system in an unauthorized manner.
In addition, Rust introduces an \textcolor{ao(english)}{\tt unsafe} keyword that allows developers to bypass compiler safety checks for certain regions of code. It is unclear if the code regions with \textcolor{ao(english)}{\tt unsafe} keywords tend to suffer from more vulnerabilities.

Several recent works perform empirical studies to characterize memory safety and concurrency bugs in Rust systems~\cite{tosem2021,pldi2020} and understand the usage of \textcolor{ao(english)}{\tt unsafe} keyword in the Rust ecosystem~\cite{icse2020evan,astrauskas2020programmers}.  
Nevertheless, the literature lacks a systematic investigation of the security risks of the Rust ecosystem. Given the popularity of Rust, a better understanding of its security risks is an important step toward sustaining and securing this software ecosystem.
To address this gap, we followed a mixed-methods approach to perform a large-scale empirical study on the vulnerabilities of the Rust ecosystem.

We compiled a dataset of 433 vulnerabilities, 300 vulnerable code repositories, and 218 vulnerability fix commits in the Rust ecosystem, spanning over 7 years in history.
With our dataset, we investigated the following research questions:

\noindent\textbf{RQ1: What are the characteristics of the vulnerabilities in the Rust ecosystem?}

Previous studies investigated the characteristics of specific types of vulnerabilities in the Rust ecosystem, e.g., memory safety~\cite{tosem2021} and concurrency issues~\cite{pldi2020}. The answer to this question aims to build a systematic view of a wide range of vulnerabilities in the Rust ecosystem, rather than a focused view of specific vulnerability types. 

In RQ1, we characterized the types, life spans, and evolution of numbers of vulnerabilities in the Rust ecosystem.
Our study identified 17 types of vulnerabilities disclosed in the Rust ecosystem, among which memory safety and concurrency issues account for two-thirds of the categorized vulnerabilities and demonstrate the fastest growth rates over time.
It takes an average of 770 days (2.1 years) for a vulnerability to be disclosed after its introduction in a code repository. One-third of the vulnerabilities have no fixes released by their public disclosure, leaving a window of opportunity for potential attacker exploitation.
The number of vulnerabilities disclosed grows slowly from November 2014 to November 2020, and has experienced two rapid growth phases, starting from November 2020 and July 2021, respectively. Meanwhile, the number of vulnerabilities introduced into code repositories demonstrates a linear growth from July 2015 to January 2020, and has stabilized since March 2020.
In addition, the normalized numbers of vulnerabilities per one thousand packages and lines of code indicate an increasing trend in package-level security risks but a decreasing trend in code-level security risks, respectively.

\noindent\textbf{RQ2: What are the characteristics of the vulnerable packages in the Rust ecosystem?}

Package reuse in software ecosystems introduces potential security risks that arise from vulnerable packages and propagate through multiple levels of dependencies among packages~\cite{kikas2017structure, zimmermann2019small}. 
A prior study~\cite{icse2020evan} investigated the usage of \textcolor{ao(english)}{\tt unsafe} keyword in Rust packages and found that the security risks of limited usage of \textcolor{ao(english)}{\tt unsafe} keyword could be amplified by their propagation through package dependencies. 
Nonetheless, it remains unclear whether unsafe code introduces more vulnerabilities compared to safe code in Rust packages. The answer to this question aims to guide practitioners in securing the Rust ecosystem. 

In RQ2, we investigated the affected versions, popularity, categorization, and affected code regions of vulnerable Rust packages.
Our study found that the vulnerable packages in the Rust ecosystem have an average of 1.3 disclosed vulnerabilities and 28.6 versions affected by the vulnerabilities. Popular packages tend to have more vulnerabilities, while unpopular packages tend to have more versions affected. 
The \emph{memory management} package category has the greatest number of vulnerabilities per package among different Rust package categories, and tends to have more \emph{memory access}, \emph{memory management}, and \emph{synchronization}  vulnerabilities as compared to other package categories.
In terms of vulnerability locality, a disclosed vulnerability affects 1.85 files, 3.35 safe functions, 0.15 unsafe functions, and 1.39 unsafe blocks on average in the vulnerable packages. 95\% of the affected functions are safe functions. Among the affected safe functions, 41.5\% contain unsafe blocks in their body.
In the vulnerable packages, vulnerable code has statistically significantly higher ratios of unsafe functions and unsafe blocks compared to complete code, implying the potential higher security risks in unsafe functions and unsafe blocks. 

\noindent\textbf{RQ3: How are the vulnerabilities in the Rust ecosystem fixed in practice?}

While numerous works have investigated general vulnerability fixes~\cite{msr2012bugfixes, SZZ, soto2016deeper, zhong2015empirical}, few have considered the vulnerability fixes in the Rust ecosystem.
The characteristics of vulnerability fixes are important to understand as they may reflect the ability to expeditiously generate fixes, verify their safety, and assess their impact on applications~\cite{CCS2017Patches}.

In RQ3, we considered the facets of vulnerability fixes such as the complexity of fixes and locality of code changes, and investigated how practitioners fix vulnerabilities in Rust packages with different localities.
The study revealed that the commits of vulnerability fixes involve an average of 41 and 18 LOC added and deleted, touching 3.85 safe functions, 0.16 unsafe functions, and 1.49 unsafe blocks on average. 96\% of the touched functions are safe functions, among which, 38.8\% contain unsafe blocks in their body.
The vulnerabilities of different types differ widely in localities of fix commits, among which, the \emph{exception management} vulnerabilities have the greatest number of safe functions touched by their fix commits. This indicates that their fixes are the least localized at the function level, indicating potential challenges of fixing in practice.

In addition, our study uncovered three patterns in the vulnerability fixes -- developers tend to (1) add safe functions or add lines in safe functions to fix vulnerable safe functions, (2) remove unsafe blocks to fix vulnerable unsafe blocks, and (3) modify unsafe trait implementations\footnote{The trait syntax in Rust is similar to the Java interface syntax.} to fix vulnerable unsafe functions.

Based on our findings, we discuss implications and provide practical lessons for securing the Rust ecosystem, such as undertaking comparable efforts into safe and unsafe code when securing Rust packages. We also highlight several research avenues, such as continuous collection and analyses of vulnerabilities to increase the awareness of security risks in the Rust ecosystem. 
This paper makes the following contributions:
\begin{itemize}
    \item We performed a large-scale empirical study to investigate the security risks in the Rust ecosystem.
    \item We provided a dataset that include 433 vulnerabilities, 300 vulnerable code repositories, and 218 vulnerability fix commits for future investigations by others\footnote{\url{https://zenodo.org/record/7828059\#.ZDo1v-xBy3Y}}.  
    \item We summarized the vulnerability fix patterns of different localities in Rust code, which can be used as guidelines to resolve vulnerabilities in practice.
    \item We provided a discussion of practical implications and outlined future avenues of research.
\end{itemize}

The replication package is online at \url{https://github.com/ZXXYy/rust_ecosystem}.

\section{Background and Preliminary Experiments}\label{sec:background}
This section gives some background on Rust, including its safety mechanisms and unsafe Rust code that are relevant to our study, and conducts preliminary experiments on Rust ecosystem.

\subsection{Rust Safety Mechanisms} 
Rust is a type-safe language designed for systems software development, which gives developers low-level control over resources but ensures memory and thread safety via a set of strict rules. The Rust compiler checks these rules to statically rule out potential safety issues. Rust programs behave like C programs, and could achieve comparable runtime performance as C programs. 
The Rust's safety mechanisms aim to prevent memory and thread safety issues that have plagued C programs. The safety mechanisms center around several basic concepts:

\begin{itemize}
\item \noindent\textbf{Ownership.} The ownership mechanism governs how a Rust program manages its memory, and prevents a Rust program from reading uninitialized memory and dangling pointers. 
Under Rust's basic ownership rule, a value (memory location)  has one exclusive owner (variable). When the owner of a value goes out of a specific scope, the value would be dropped or freed. 
The variable assignment leads to the transfer of ownership. Once a variable loses the ownership of a value, the variable would become unusable. 
\item \noindent\textbf{Borrowing.} To enable sharing a value without moving its ownership, the borrowing mechanism allows the creation of a reference and passes the reference to another variable. In addition, Rust supports multiple shared immutable references, i.e., references that allow read-only aliasing. Rust enforces the memory locations reachable by a shared reference to be immutable to prevent data races and inadvertent side effects.
\item \noindent\textbf{Lifetime.} Lifetime explains the scopes for which references in a Rust program are valid. 
The lifetime feature in Rust includes a variety of generics that indicate how references relate to each other. 
Specifically, to determine when references go out of their scopes, the compiler associates each borrowed reference with a lifetime and tracks constraints between references. The lifetime inference assures that the lifetime of a borrowed ownership would last long enough for use.
\end{itemize}

\subsection{Unsafe Rust Code}
Rust developers usually need flexibility in writing their code, including accessing arbitrary memory with C-style pointers, invoking system calls, and accessing global static memory. Rust allows programs to bypass its security mechanisms with the \textcolor{ao(english)}{\tt unsafe} keyword. Code regions marked with the \textcolor{ao(english)}{\tt unsafe} keyword could bypass Rust's compiler checks, and be able to perform five types of operations: dereferencing and manipulating raw pointers, calling unsafe functions, accessing or modifying mutable static variables (i.e., global variables), implementing unsafe traits, and accessing fields of {\tt unions}. 
For simplicity, we use the phrase ``unsafe code'' throughout the paper to refer to the code regions that are marked with the \textcolor{ao(english)}{\tt unsafe} keyword.
The code regions that can be marked as unsafe include: 
\begin{itemize}
\item \noindent\textbf{Unsafe Blocks.} 
An unsafe block defines a block of Rust code in which some compiler safety checks would be disabled. For instance, as shown in \autoref{lst:unsafe_block_example}, an unsafe block  dereferences a raw pointer {\tt r}. The dereferencing operation can bypass compiler checks due to the \textcolor{ao(english)}{\tt unsafe} keyword.
Note that if a block of Rust code is marked with the  \textcolor{ao(english)}{\tt unsafe} keyword but does not contain any of the aforementioned five types of operations, the compiler would emit a warning message.

\begin{minipage}{\linewidth}
\begin{lstlisting}[caption={Unsafe block example.},label={lst:unsafe_block_example},language=Rust,basicstyle=\small\ttfamily]
let mut address = 5;
// create a raw pointer
let r = &address as *const i32;
unsafe {
    print!("r is: {}", *r)
}
\end{lstlisting}
\end{minipage}

\item\noindent\textbf{Unsafe Functions.} 
A Rust function can be declared as an unsafe function with the \textcolor{ao(english)}{\tt unsafe} keyword. The \textcolor{ao(english)}{\tt unsafe} keyword requires the callers of unsafe functions to satisfy some preconditions or bypass compiler checks via unsafe blocks.
If an unsafe function only includes safe operations, the compiler would not emit a warning because it cannot tell whether programmers do it intentionally or by mistake. \autoref{lst:unsafe_function_example} shows a typical usage of an unsafe function: the unsafe function {\tt bar()} is called within an unsafe block of the safe function {\tt foo()}, indicating that unsafe functions are encapsulated by their callers. 

\begin{minipage}{\linewidth}
\begin{lstlisting}[caption={Unsafe function example.},label={lst:unsafe_function_example},language=Rust,basicstyle=\small\ttfamily]
unsafe fn bar() {...}

fn foo() { // a safe function
    unsafe {
        bar(); // call an unsafe function in an unsafe block
    } 
}
\end{lstlisting}
\end{minipage}
\par

\item \noindent\textbf{Unsafe Traits}. 
The trait is an advanced feature in the Rust type system to enable inheritance. In general, traits of Rust are similar to interfaces to Java or abstract classes to C++. 
A trait can be declared as unsafe with the  \textcolor{ao(english)}{\tt unsafe} keyword if it contains unsafe functions or its implementations is required to satisfy any invariant.
\end{itemize}

\subsection{Preliminary Investigation of Rust Ecosystem}
\begin{figure}[t]
\centering
    \begin{minipage}{0.48\textwidth}
        \centering
        \includegraphics[width=6cm]{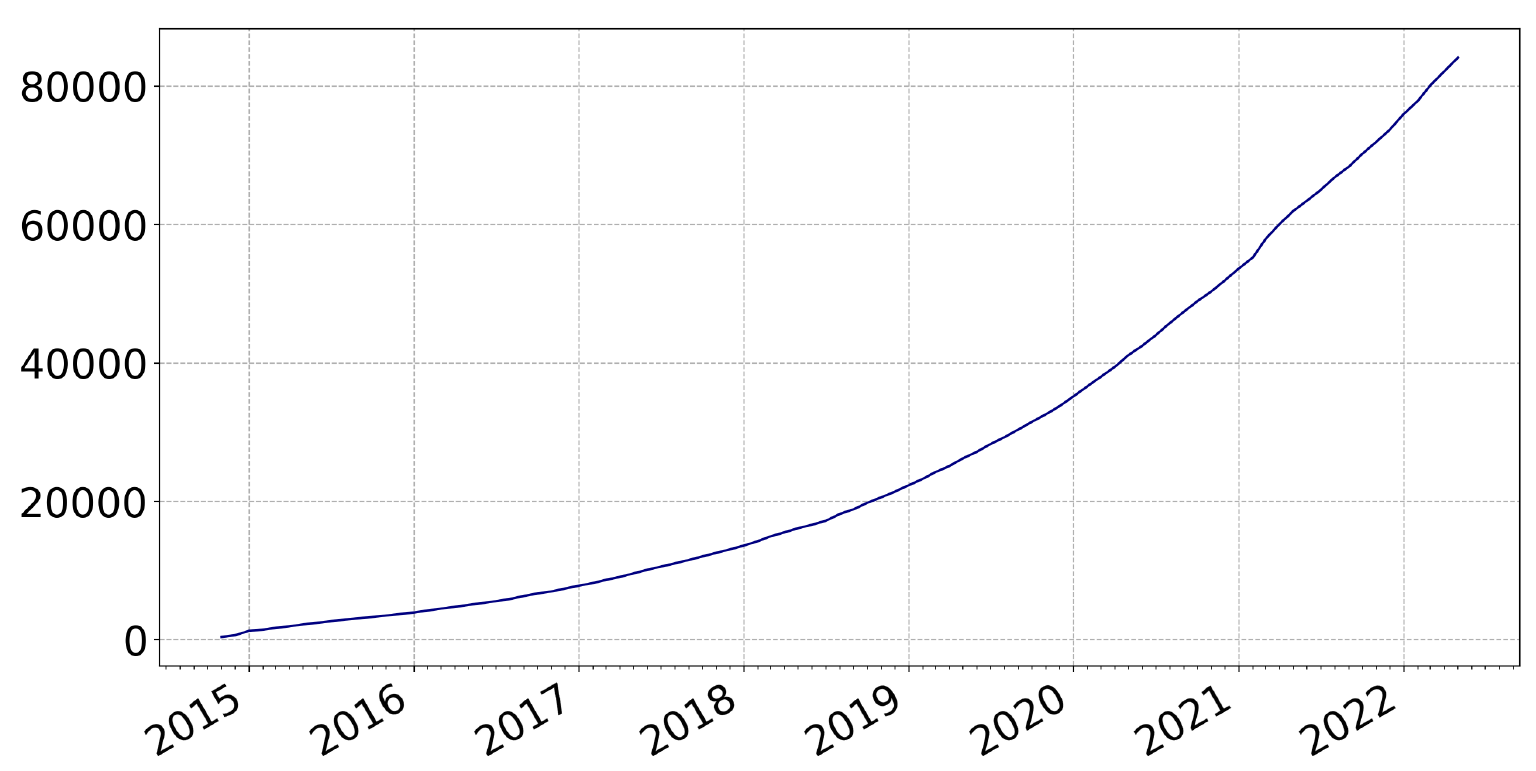}
\caption{Evolution of number of packages per month.}
        \label{fig:numlib}
    \end{minipage}
    \begin{minipage}{0.48\textwidth}
        \centering
        \includegraphics[width=6cm]{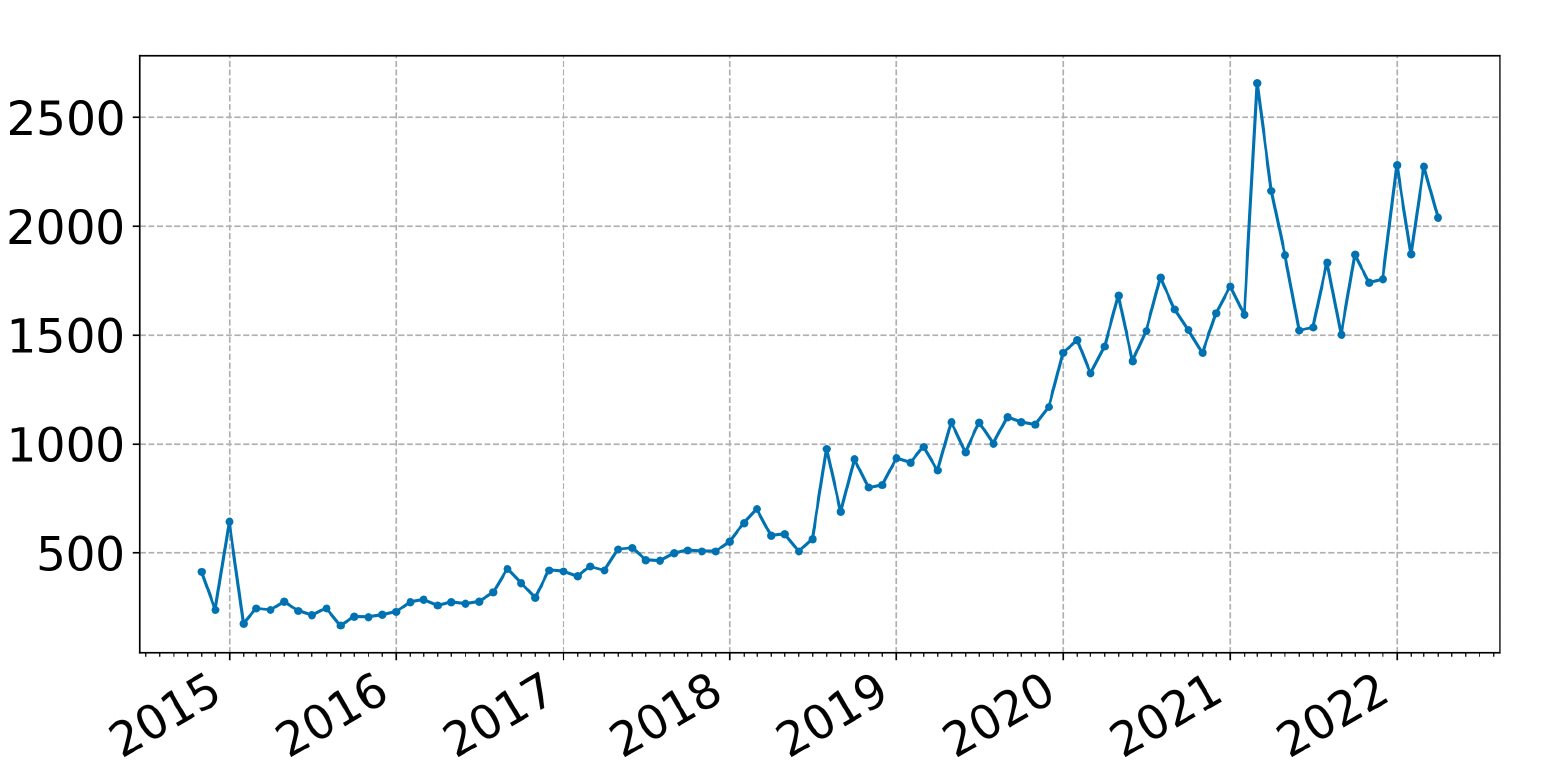}
        \caption{Growth rate of packages created per month.}
        \label{fig:numCreate}
    \end{minipage}
\end{figure}

Rust is a striving ecosystem with ongoing and even accelerating growth in the number of packages and downloads. An increasing number of areas start to choose Rust as the programming language for software development. \autoref{fig:numlib} shows the evolution of the number of packages in the Rust ecosystem since its inception. The first package in the Rust ecosystem hosted on {\tt crates.io} was published on November 11, 2014. The number of packages grows 1.6x per year on average from 2015 to 2020. From 2020 to 2021, the growth in the number of packages slightly slows down, exhibiting a 1.4x growth rate. 
\autoref{fig:numCreate} shows the number of packages that are being created on {\tt crates.io} every month since November 2014. From late 2014 to early 2018, the increasing number of packages per month is less than 500. Since early 2018, the growth in the increasing number of packages per month accelerates, experiencing a peak in March 2021, which may be due to the official announcement of the Rust Foundation\footnote{\url{https://foundation.rust-lang.org/news/2021-02-08-hello-world/}} on February 8, 2021. Following the peak, the increasing number of packages per month shows steady growth and resembles the trend prior to 2021, indicating that the announcement acted as a boost for the Rust ecosystem.
\begin{figure}[t]
    \centering
    \includegraphics[width=0.48\linewidth]{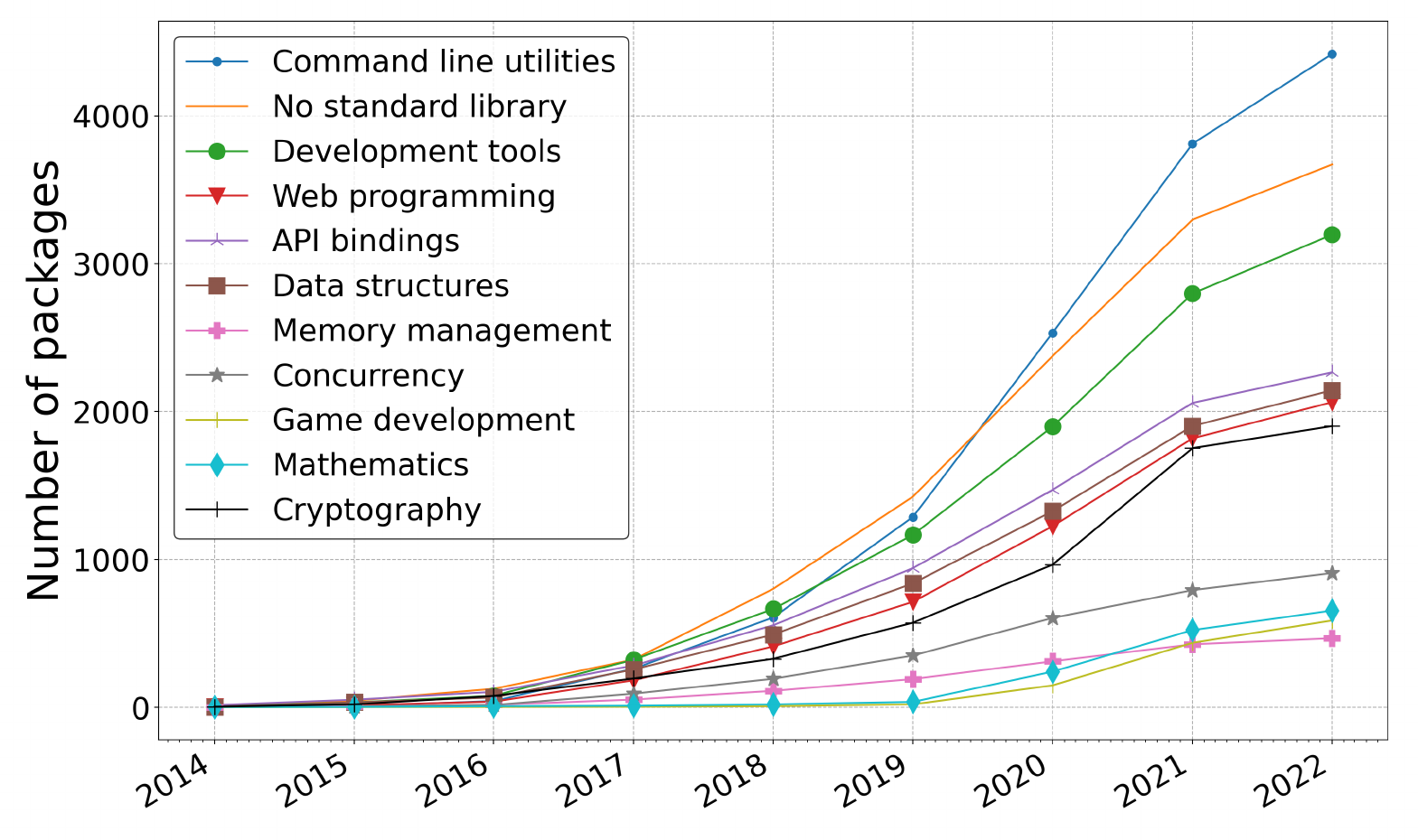}
    \caption{Evolution of numbers of packages per year across package categories.}
    \label{fig:numlib_across_cat}
\end{figure}
\begin{figure}[t]
\centering
        \begin{minipage}{0.48\textwidth}
        \centering
        \includegraphics[width=\linewidth]{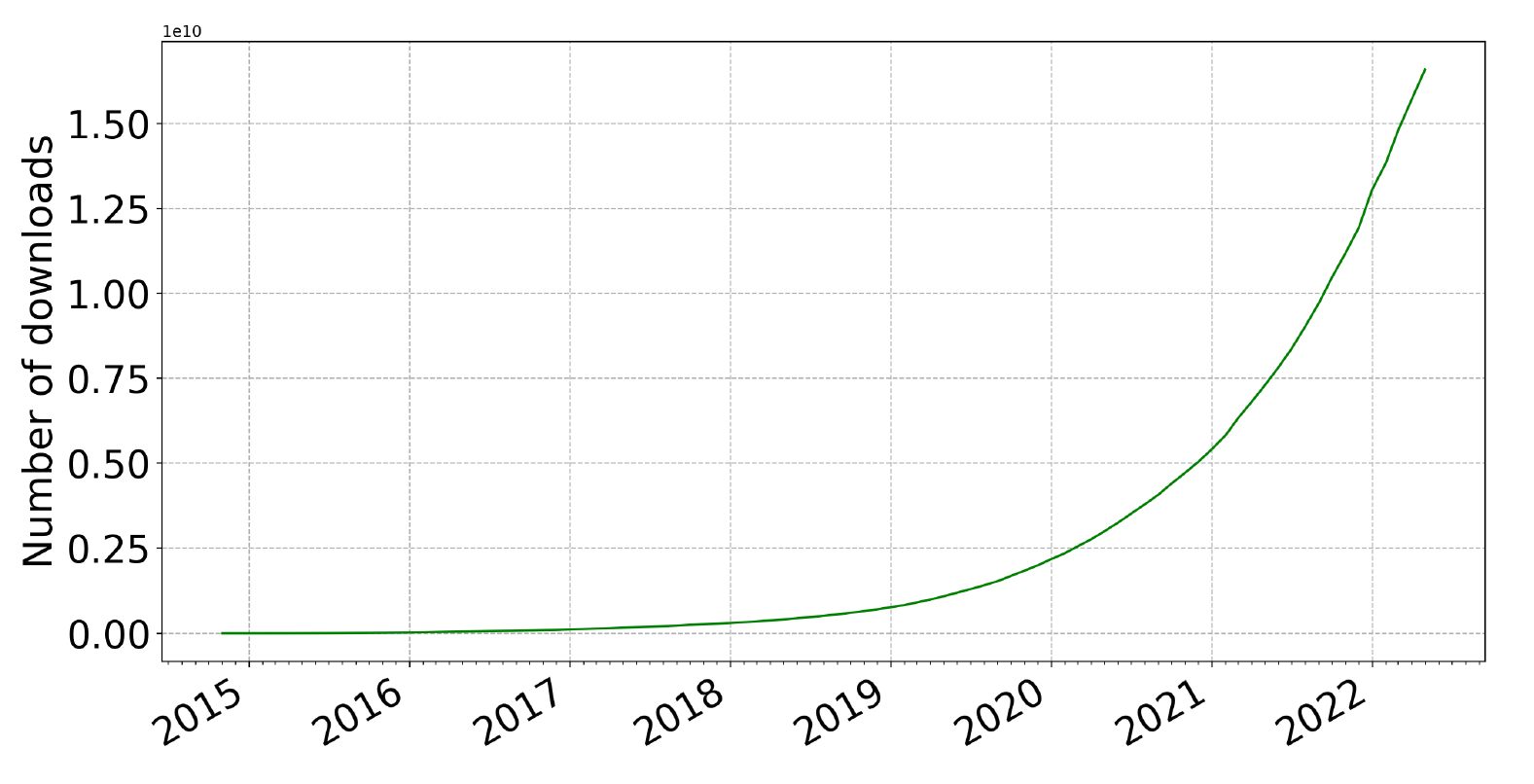}
        \caption{Evolution of package downloads.}\label{fig:downloadEvolution}
    \end{minipage}
    \begin{minipage}{0.48\textwidth}
        \centering
        \includegraphics[width=6cm]{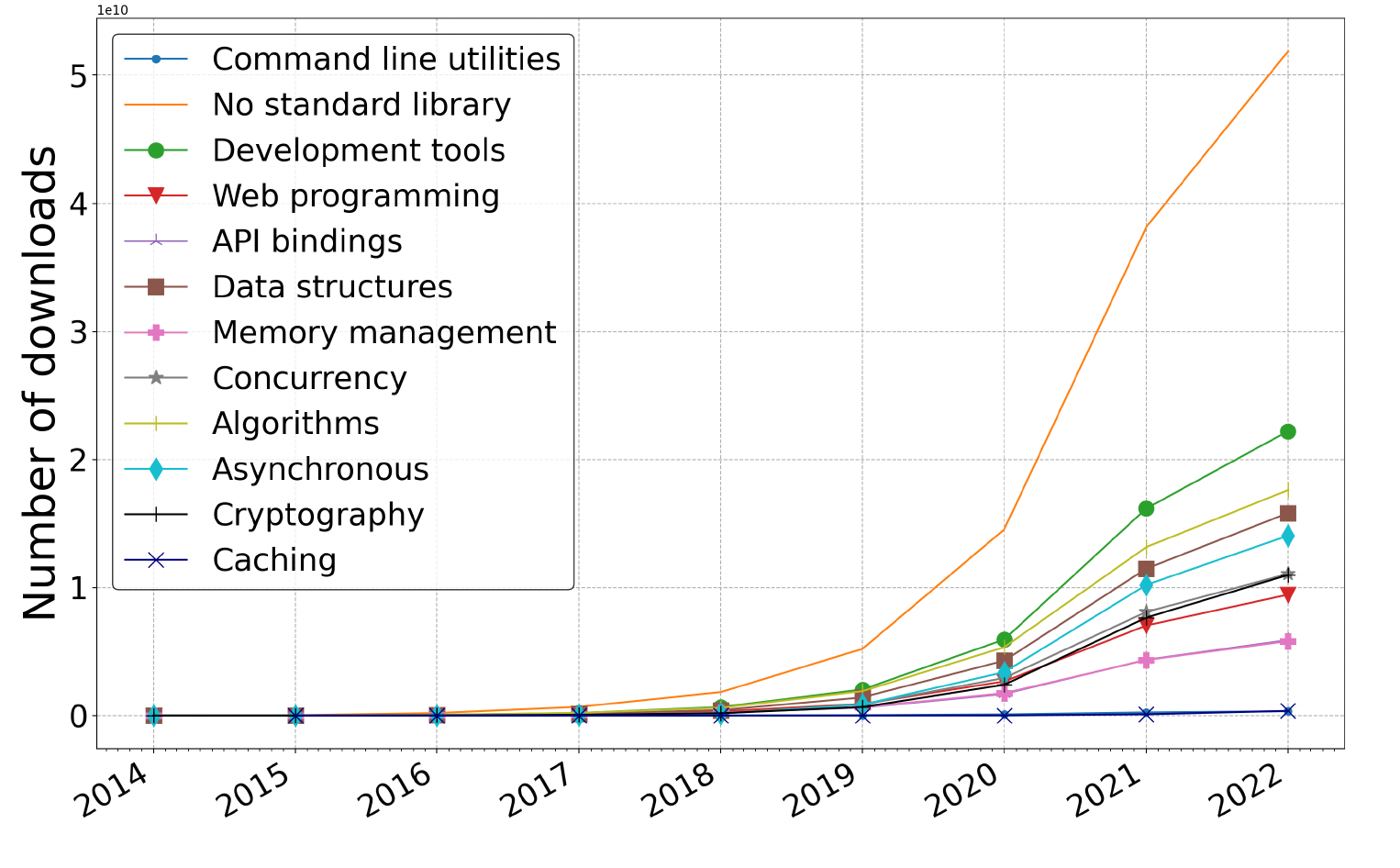}
        \caption{Evolution of package downloads across package categories.} \label{fig:numDownload_perCat}
     \end{minipage}
\end{figure}
As shown in \autoref{fig:numlib_across_cat}, the top 5 categories with the most packages, \emph{command line utilities}, \emph{no standard library}, \emph{development tools}, \emph{api bindings} and \emph{data structures}, undergo continuous near-exponential growth in the number of packages over time, which resembles the growth of the Rust ecosystem.

We also investigate the downloads of all Rust packages on a monthly basis from November 2014 to May 2022. As \autoref{fig:downloadEvolution} shows, the downloads of Rust packages grow exponentially from November 2014 to May 2022. The growth rate of downloads has experienced  a dramatic increase since April 2019, which is far greater than the growth rate of package numbers.
\autoref{fig:numDownload_perCat} presents the top 5 categories of Rust packages that have the greatest downloads of packages, i.e., \emph{no standard library}, \emph{development tools}, \emph{algorithms}, \emph{data structures} and \emph{asynchronous}. The 5 categories of Rust packages experience a continuous near-exponential growth in downloads over time, resembling the growth of package downloads in the Rust ecosystem. The number of packages in the \emph{memory management} and \emph{concurrency} categories grow linearly as indicated by regression analysis ($R^2=0.9163$ and $0.9460$).

\section{Research Methodology}\label{sec:methodology}
\begin{figure*}[t]
  \centering
  \includegraphics[width=0.95\linewidth]{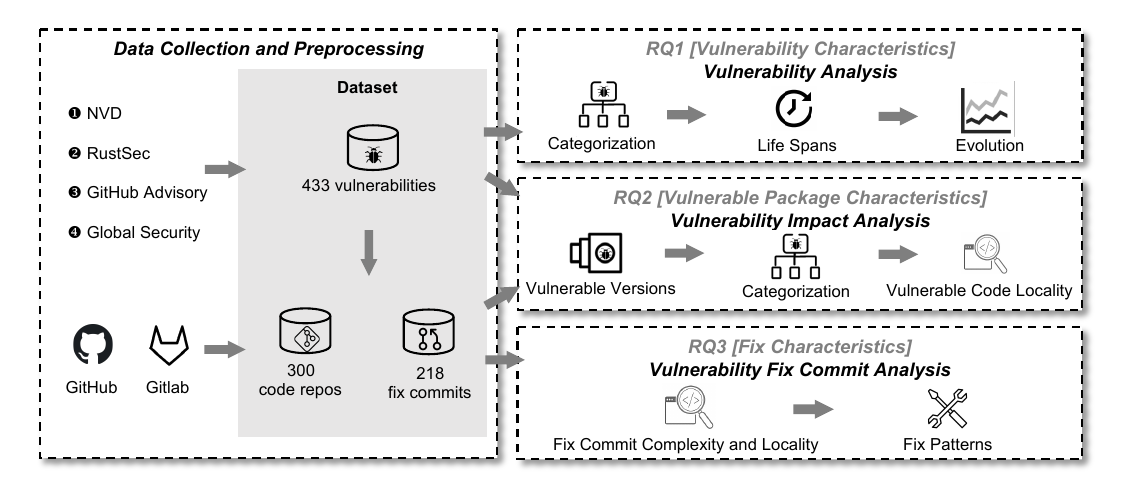}
\caption{Overview of research methodology.}
\label{fig:dataflow}
\end{figure*}

We designed and conducted a mixed-methods empirical study, analyzing a dataset of vulnerabilities, vulnerable packages, and vulnerability fixes in the Rust ecosystem, as depicted in \autoref{fig:dataflow}. Our research methodology is detailed in the following subsections.

\subsection{Data Collection and Preprocessing}

\noindent\textbf{Step 1: Collecting vulnerabilities in the Rust ecosystem.}
We collected an initial set of 776 vulnerabilities disclosed on OSV\footnote{\url{https://osv.dev}} from November 11, 2014, to May 24, 2022. OSV is a distributed vulnerability database for open source ecosystems, which serves as an aggregator of vulnerability databases including GitHub Security Advisories, RustSec, the National Vulnerability Database (NVD), and the Global Security Database. We identified 343 duplicated vulnerabilities from the initial set  and merged the information of vulnerabilities with the same references to NVD or RustSec, resulting in a final set of 433 unique vulnerabilities.
For each vulnerability, we obtained the summary, detail, published date, modified date, vulnerability introduced and fixed versions, references (e.g., code repository, fix commit, issue, and pull request), and type(s) of the vulnerability. 

\noindent\textbf{Step 2: Locating code repositories of vulnerable Rust packages.}
We further located the code repositories of vulnerable Rust packages by following the references provided by disclosed vulnerabilities. 
We found that 17 disclosed vulnerabilities
from 13 Rust packages do not provide references to their code repositories, thus we did not consider the 17 vulnerabilities in this step. As a result, we obtained a total of 300 code repositories for vulnerable packages on GitHub and GitLab.

\noindent\textbf{Step 3: Identifying vulnerability-fix commits in vulnerable code repositories.} 
We identified an initial set of 287 vulnerability-fix commits by analyzing three types of fix  references that are provided by collected vulnerability reports: 
\begin{itemize}
    \item \emph{Commit.} We considered the commit in the reference as the vulnerability-fix commit.
    \item \emph{Pull Request.} Given a pull request could have multiple commits, we identified vulnerability-fix commits by searching vulnerability-fix related keywords in commit messages, including fix, repair, error, bug, issue, exception and cve.
    \item \emph{Issue.} We located the pull requests or commits related to an issue to identify the vulnerability-fix commits. We only considered the closed issues because they indicate that the corresponding vulnerabilities are fixed. 
\end{itemize}

With the initial set of vulnerability-fix commits, we further excluded commits that do not have code for vulnerability fixing, .e.g., for refactoring purpose. Specifically, we inspected the vulnerability-fix commits in the initial set and excluded 69 commits irrelevant to vulnerability fixes, including 34 commits that modified change logs, corrected spelling or styling, 28 that reported the packages as unmaintained and did not fix any vulnerabilities, and 7 that involved code refactoring in the fix. As a result, we collected 218 vulnerability-fix commits for 180 vulnerabilities in the Rust ecosystem.

\subsection{Processing Vulnerability Fixes}

\begin{figure}[t]
  \centering
  \includegraphics[width=0.7\linewidth]{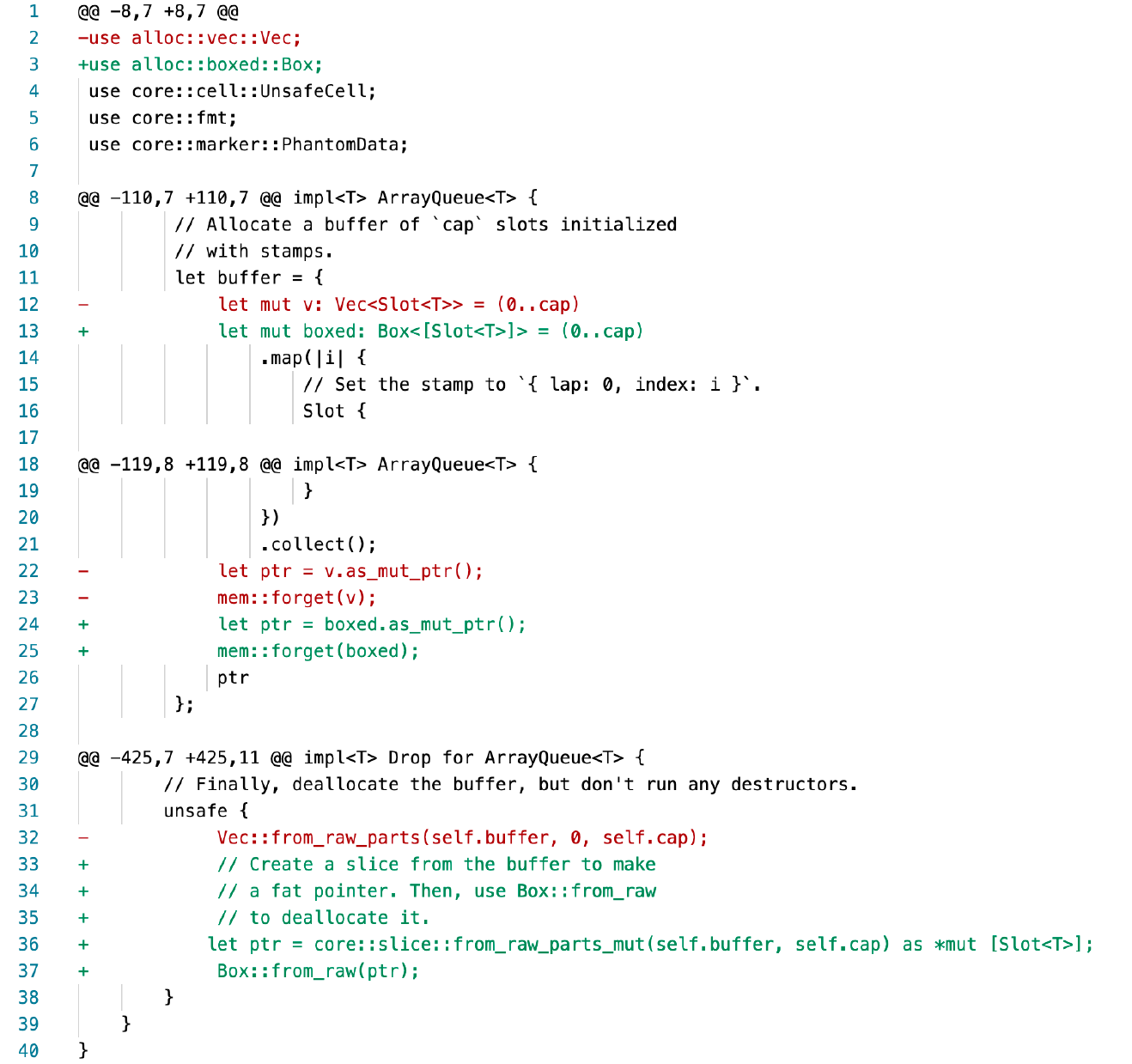}
\caption{An example fix commit applied to vulnerability \emph{RUSTSEC-2020-0052} of {\tt crossbeam-channel}.}
  \label{fig:diff_example}
\end{figure}

For each fix commit we collected, we considered the removed lines in the fix commit as \emph{vulnerable code}, and the added lines as \emph{fixing code}, as with prior work~\cite{SZZ,tufano2018empirical,ray2016naturalness,chen2021neural}. Similar to prior work~\cite{CCS2017Patches}, we excluded non-source code files, e.g., documentation, change logs, and test files, and further removed non-functional source code, e.g., empty lines, comment lines, and lines that are not inside any functions. We used git diff wrapped in PyDriller~\cite{pydriller} to obtain the textual diffs of the commit and located the removed and added lines.
\autoref{fig:diff_example} gives an example of a fix commit that represents code changes as textual diffs. The black, \textcolor{red}{red} and \textcolor{ao(english)}{green} colors represent unchanged code, deleted lines, and added lines, respectively. In the example fix commit, lines 12, 22, 23, and 32 are vulnerable code; lines 13, 24, 25, and 36 are fixing code; and lines 2, 3, 33, 34, and 35 are non-functional source code. 

We further developed a Rust compiler plugin to determine whether the functions and blocks in the vulnerability-fix commits marked \textcolor{ao(english)}{\tt unsafe}.
First, we extracted the two versions of affected files before and after the fix commit. 
Second, we extracted the line numbers of vulnerable and fixing code for each affected file $u$ (i.e., compilation unit) before and after the commit, namely, $LineVul^u$ and $LineFix^u$. We denote the code ranges of a function and an unsafe block in the compilation unit $u$ by $FRange$ and $UBRange$, respectively. 
Third, the plugin identified the code ranges of each function in the affected files before and after the commit, namely, $(FRange,  FRange^{'})$, as well as the code ranges of each \textcolor{ao(english)}{\tt unsafe} block, namely, $(UBRange, UBRange^{'})$. 
Finally, to locate vulnerable code with respect to functions and unsafe blocks in a compilation unit, we checked whether the range of vulnerable code is inside $FRange$ and $UBRange$. 
Specifically, for each function $f$ in a compilation unit $u$, we checked whether there exists $LineVul^{u}$ in the corresponding affected file, such that $LineVul^{u} \cap FRange_{f} \neq \emptyset$. If so, we considered the function $f$ as a vulnerable function. Similarly, we identified vulnerable \textcolor{ao(english)}{\tt unsafe} block $b$, such that $LineVul^{u} \cap UBRange_{b} \neq \emptyset$. 
Likewise, to locate fixing code with respect to functions and unsafe blocks in a compilation unit, the plugin checked whether the range of fixing code is inside $FRange'$ and $UBRange'$.
Specifically, for each function $f$ in a compilation unit $u$, we checked whether there exists $LineFix^{u}$ in the corresponding affected file, such that $LineFix^{u} \cap FRange_{f}^{'} \neq \emptyset$. If so, we considered the function $f$ as a fixing function. Similarly, we identified fixing \textcolor{ao(english)}{\tt unsafe} block $b$, such that $LineFix^{u} \cap UBRange_{b}^{'} \neq \emptyset$. 

\subsection{Characterizing Vulnerabilities, Vulnerable Packages, and Fixes}

\begin{table*}[t]
\caption{Vulnerability types with mappings between classification schemes.}
  \label{tab:classification_scheme}
\footnotesize
\begin{tabular}{p{3cm}p{7cm}c}
\toprule
    \textbf{Vulnerability Type} & \textbf{CWE ID} 
    & \textbf{RustSec Category}\\
    \midrule
    Memory Access           & 118, 119, 120, 121, 122, 125, 126, 127, 131, 135, 170, 416, 467, 476, 588, 785, 787, 824
     & memory-exposure \\
    Memory Management       & 415, 590, 761, 762, 763    
    & memory-corruption \\
    Synchronization         & 362, 363, 364, 366, 367, 370, 412, 413, 414, 543, 567, 585, 609, 638, 662, 667, 764, 765      
    & thread-safety \\
    Tainted Input           & 15, 20, 74, 77, 78, 643, 644, 652, 687, 129     
    & format-injection \\
    Resource Management    & 400, 404, 459, 672, 674, 770, 774, 772, 789        
    & denial-of-service \\
    Exception Management    & 248, 252, 253, 273, 280, 390, 431, 478, 484, 584, 600, 665, 908, 909       
    & - \\
    Cryptography            & 327, 347, 1240        
    & cryptography \\
Risky Values            & 28, 190, 194, 369, 456, 466, 468, 475, 480, 486, 562, 570, 579, 587, 594, 597, 681, 685, 704, 768, 843        
    & - \\
    Path Resolution         & 22, 30, 42, 51, 57, 58, 59, 62, 64, 65, 67, 73, 243, 706        
    & file-disclosure \\
    Information Leak        & 8, 14, 117, 200, 214, 226, 244, 256, 311, 374, 403, 495, 501, 523, 532, 591, 598, 607, 642, 668, 767       
    & - \\
    Privilege               & 269, 272         
    & privilege-escalation \\
    Predictability          & 330, 338, 340         
    & - \\
    Authentication          & 259, 293, 306, 307, 321, 350, 360, 422, 425, 565, 605, 620, 295
         & - \\
    API                     & 111,242,245,382,474,477,479,558,572,586,589,617,676,758        
    & - \\
    Access Control          & 279, 285, 424         
    & - \\
    Failure to Release Memory & 401       
    & - \\
        Other                   & 188, 193, 657, 670, 682, 697, 835      
    & code-execution \\
    \bottomrule
  \end{tabular}
\end{table*}

\noindent\textbf{Vulnerability categorization  (RQ1).}
The vulnerabilities from the four sources use two classification schemes, i.e., Common Weakness Enumeration (CWE) and RustSec categorization. To categorize disclosed vulnerabilities in the Rust ecosystem, we leveraged Software Fault Patterns (SFP)~\cite{SFP} to build connections between two classification schemes of vulnerabilities. As shown in \autoref{tab:classification_scheme}, we identified 17 vulnerability types in our dataset (the \emph{Vulnerability Type} column), with the corresponding CWE IDs and RustSec categorization in the \emph{CWE ID} and \emph{RustSec Category} columns. We further categorized the disclosed vulnerabilities in our dataset into the 17 vulnerability types.

\noindent\textbf{Vulnerability life spans (RQ1).} Upon a vulnerability's disclosure, we might ask how long it plagued a code repository before a developer fix the vulnerability. We denote the duration as the \emph{life span} of the vulnerability in the code repository, which is investigated in prior work~\cite{CCS2017Patches}. In the life span of a vulnerability, we also measured two duration: (1) the duration between the introduction and disclosure of a vulnerability, which reflects the window of opportunity for attackers who silently discover a vulnerability to leverage it offensively, before any defensive measures are taken, and (2) the duration between the disclosure and fixing of a vulnerability, which affects the the remediation process and the potential impact of the vulnerability.
Reliably determining when a vulnerability was born and fixed automatically is challenging, as it requires understanding the source code and the nature of the vulnerability~\cite{CCS2017Patches}. Thus, we utilized the collected and processed vulnerability-fix commits. Specifically, for all lines of code deleted by a fix commit, i.e. $LineVul$, we used \emph{git blame} to retrieve the last modification date of each line~\cite{SZZ}. Note that we ignore the commits with only additions due to newly added lines did not exist prior to the commit. We conservatively designate the earliest blame date across all lines as the estimated date of vulnerability introduction. We used the commit date of the fix commit to estimate when the vulnerability is fixed. In case the vulnerability had multiple fix commits, we conservatively designated the most recent commit date of the fix commits as the estimated date of vulnerability fix. 

\noindent\textbf{Vulnerability evolution (RQ1).} 
We investigated the evolution of numbers of vulnerabilities that are introduced and disclosed in the Rust ecosystem over time. To mitigate the impact from the package growth on vulnerability evolution, we normalized the number of disclosed vulnerabilities to both the number of packages and the lines of code (LOC) of vulnerable packages in the ecosystem. In addition, we compared the evolution of numbers of disclosed vulnerabilities across vulnerability types and package categories. 

\noindent\textbf{Affected versions of vulnerable packages (RQ2).} We extracted package names from the vulnerability reports and identified the unique set of packages that contained at least one vulnerability. For each vulnerable package, we aggregated the affected versions by each of its corresponding vulnerabilities as indicated in the vulnerability reports. We also analyzed the packages with the most vulnerabilities to investigate whether the package popularity would impact the number of vulnerabilities disclosed in a package. 

\noindent\textbf{Vulnerable package categorization (RQ2).} We categorized the vulnerable packages by referring to the categorization information provided by {\tt crates.io}. Given {\tt crates.io} does not provide categorization information for 165 vulnerable packages, we identified an average of 2.10 package categories (median = 2) for the rest 172 vulnerable packages with categorization information. We further compared the numbers of vulnerabilities, total packages and downloads, as well as the distributions of vulnerability types across different package categories. 

\noindent\textbf{Vulnerable code locality (RQ2).} For each vulnerability, we first counted the numbers of files, functions, and unsafe blocks that are touched by the corresponding vulnerable code. We then used the total numbers of functions and unsafe blocks in affected versions of vulnerable packages (i.e., the version before vulnerability fixes) as the baseline for normalization. For code that failed in compilation, we used regular expression to estimate the numbers of functions and unsafe blocks. 
The resulting baseline is shown in \autoref{tab:total_loc}: the vulnerable packages contain an average of 637.22 safe functions, 16.69 unsafe functions and 77.30 unsafe blocks (a median of 226 safe functions, 2 unsafe functions, and 14 unsafe blocks).
With the baseline, we further measured the ratios of unsafe functions and unsafe blocks touched by vulnerable code in vulnerable packages, and compared them with the corresponding ratios in the complete code of vulnerable packages.
In addition, we compared the vulnerable code localities across different vulnerability types in terms of numbers of commits, files, safe and unsafe functions, and unsafe blocks.

\begin{table}[t]
  \caption{Descriptive statistics of vulnerable packages.}
  \label{tab:total_loc}
  \footnotesize
  \begin{tabular}{lcccc}
    \toprule
& \textbf{\# Safe Functions} & \textbf{\# Unsafe Functions} & \textbf{\# Unsafe Blocks} \\
    \midrule
    mean ($\mu$)    & 637.22    & 16.69     & 77.30\\
    median (M)      & 226       & 2         & 14\\
    min             & 2         & 0         & 0\\
    max             & 7,804     & 361       & 1,804\\
    std             & 1,176.03   & 49.07    & 181.58\\
    total           & 137,003    & 3,589      & 16,620  \\
    \bottomrule
  \end{tabular}
\end{table}

\noindent\textbf{Fix commit complexity and locality (RQ3).} To investigate the complexity of a fix commit, we used lines of code (LOC) touched by the fix commit, .i.e, its vulnerable and fixing code, as a simple-albeit-rudimentary metric as with prior studies~\cite{ccs2017-18,ccs2017-26,SZZ,ccs2017-41}.
Meanwhile, to investigate the locality of a fix commit, we first counted the numbers of functions, unsafe functions, and unsafe blocks touched by its vulnerable and fixing code.
In addition, we compared the localities of vulnerability fix commits across vulnerability types in terms of numbers of commits, files, safe and unsafe functions, and unsafe blocks.

\noindent\textbf{Fix patterns (RQ3).} We inspected vulnerability fix commits and summarized fix patterns in the fix commits with identical locality category. Each fix commit could fall into multiple locality categories from three categories: (1) the \emph{safe function} category, if the vulnerable code in the fix commit includes safe function(s), (2) the \emph{unsafe function} category, if its vulnerable code includes unsafe function(s), and (3) the \emph{unsafe block} category, if its vulnerable code includes unsafe block(s).

\section{Results}\label{sec:result}
In this section, we present the results of our research questions that investigate the security risks in the Rust ecosystem. 

\subsection{RQ1: Vulnerabilities in the Rust Ecosystem}
We investigated the characteristics of disclosed vulnerabilities in the Rust ecosystem, including the vulnerability types, life spans, and the evolution of the number of vulnerabilities.

\begin{table}[t]
  \caption{Distribution of vulnerabilities across vulnerability types. "Percentage" denotes the number of vulnerabilities belonging to a specific vulnerability type divided by the number of categorized vulnerabilities, .i.e, count/360. "Disclosure Duration" denotes the median duration between introduction and disclosure reported in days. "Fix Duration" denotes the median duration between disclosure and fix reported in days. }
\label{tab:cve_type}
  \small
  \begin{tabular}{lrlrp{1.5cm}l}
    \toprule
\textbf{Vulnerability Type} & \textbf{Count} & \textbf{(with Fix)}& \textbf{Percentage} & 
    \textbf{Disclosure Duration} & \textbf{Fix Duration} \\
\midrule
    Memory Management       & 144  & (67)    & 40.00\% & 668.0  &1.0\\
    Memory Access           & 141  & (53)    & 39.17\% & 678.0  & 0.0\\
    Synchronization         & 74   & (44)    & 20.56\% & 770.5  & 0.0\\
    Tainted Input           & 46   & (23)    & 12.78\% & 780.0  & -5.0\\
    Resource Management     & 40   & (37)    & 11.11\% & 599.0  & -2.0\\
    Exception Management    & 38   & (18)    & 10.56\% & 1062.5 & 18.0\\
    Cryptography            & 26   & (8)     & 7.22\%  & 757.5  & -2.0\\
    Other                   & 26   & (10)    & 7.22\%  & 419.5  & -1.0\\
    Risky Values            & 21   & (10)    & 5.83\%  & 802.0  & -2.0\\
    Path Resolution         & 14   & (12)    & 3.89\%  & 165.5  & -2.0\\
    Information Leak        & 9    & (3)     & 2.50\%  & 80.0   & -14.0\\
    Privilege               & 4    & (1)     & 1.11\%  & 76.0   & -5.0\\
    Predictability          & 3    & (2)     & 0.83\%  & 107.5  & -2.0\\
    Authentication          & 3    & (0)     & 0.83\%  & /      & / \\
    API                     & 2    & (1)     & 0.56\%  & 846    & 30.0\\
    Access Control          & 2    & (0)     & 0.56\%  & /      & / \\
    Failure to Release Memory & 1  & (0)     & 0.28\%  & /      & /\\
    \bottomrule
  \end{tabular}
\end{table}

\noindent\textbf{Types of vulnerabilities.} We collected a total of 433 unique vulnerabilities in the Rust ecosystem, out of which 73 have not been categorized, leaving 360 vulnerabilities that are categorized with a median of 1 vulnerability type (min: 1, max: 4, mean: 1.65, std: 0.76). \autoref{tab:cve_type} presents the overall distribution of vulnerabilities across 17 vulnerability types. Memory safety and concurrency issues account for 63.6\% of the 360 categorized vulnerabilities.

Memory safety issues involve \emph{memory access} (39.17\%) and \emph{memory management} (40.00\%) vulnerability types, accounting for 59.7\% of the categorized vulnerabilities. 
The \emph{memory access} vulnerabilities usually arise from buffer or pointer access problems, e.g., \textit{buffer overflow}, \textit{use after free}, and \textit{null pointer deference}. We take the RUSTSEC-2021-0128 in the {\tt rusqlite} package as an example of \emph{memory access} vulnerability. In the vulnerable code of the {\tt rusqlite} package affected by RUSTSEC-2021-0128, the lifetime bounds on several closure-accepting functions are so loose that allow the access to dropped objects on the stack thus cause \textit{use after free} error.
The \emph{memory management} vulnerabilities are due to problems in memory allocation or deallocation, e.g., \textit{double free}. RUSTSEC-2021-0033 in the {\tt stack\_dst} package is an example of \emph{memory management} vulnerabilities. Specifically, the {\tt push\_cloned} function in the vulnerable code of {\tt stack\_dst} package deallocates uninitialized memory thus cause \textit{double free} error.

Concurrency issues involve \emph{synchronization} vulnerabilities, which rank the third in the frequency of occurrence across different types of vulnerabilities (20.56\%). 
The \emph{Synchronization} vulnerabilities occur when multiple processes or threads share resources, including race condition and misuse of locks. For instance, the unsafe {\tt Send} trait implementation in the {\tt atom} package involved in RUSTSEC-2020-0044 causes data race error.

\begin{figure}[t]
\centering
    \begin{minipage}{0.48\textwidth}
        \centering
        \includegraphics[width=6.5cm]{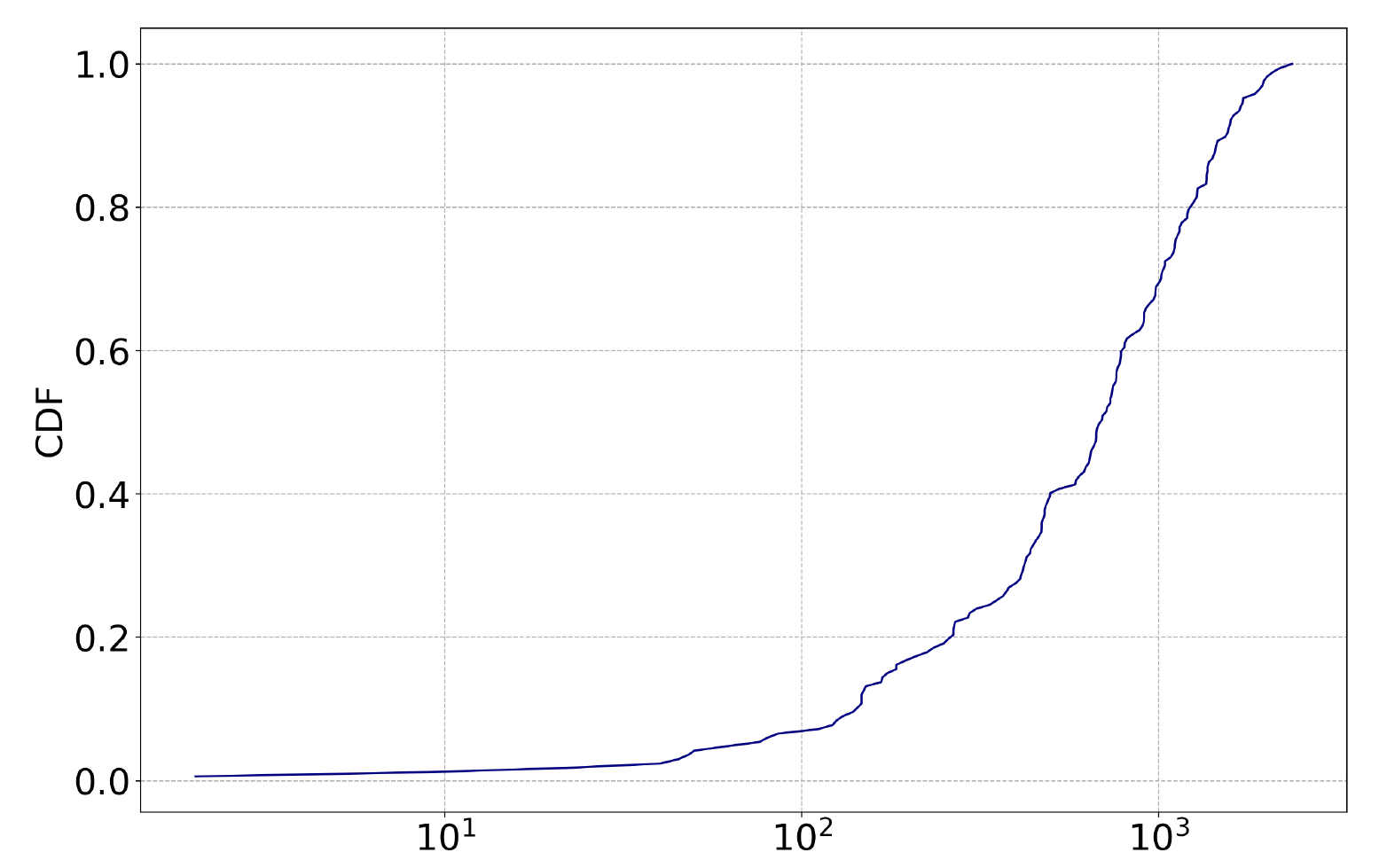}
        \caption{CDFs of the duration between the introduction and disclosure of a vulnerability.}
        \label{fig:vul_disclose_duration}
    \end{minipage}
    \begin{minipage}{0.48\textwidth}
        \centering
        \includegraphics[width=6.5cm]{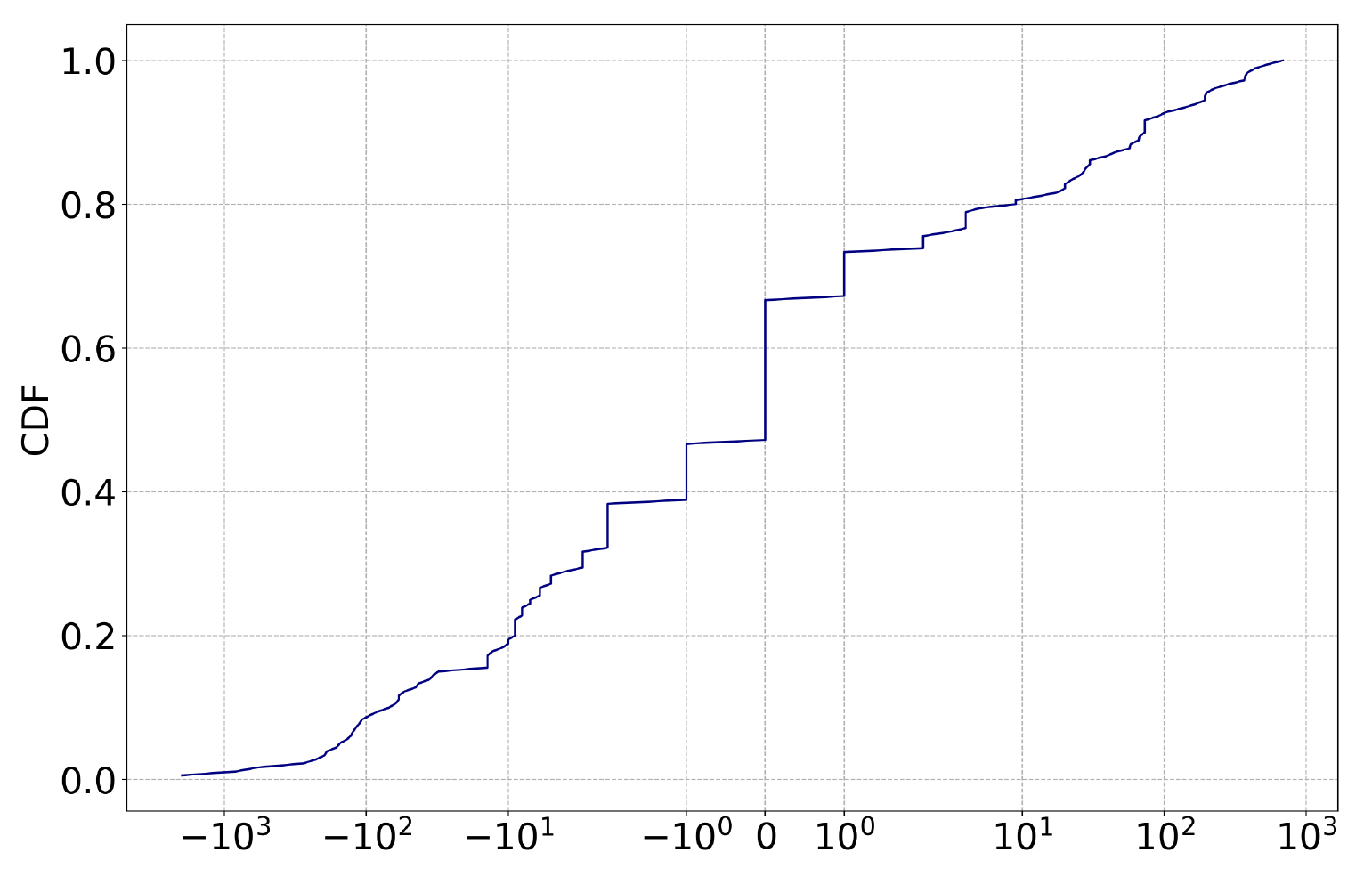}
        \caption{CDFs of the duration between the disclosure and fixing of a vulnerability.}
        \label{fig:vul_fix_duration}
    \end{minipage}
    
\end{figure}

\noindent\textbf{Vulnerability life spans.}
\autoref{fig:vul_disclose_duration} illustrates the distribution of disclosure duration for vulnerabilities. It takes an average of 770 days (2.1 years) for a vulnerability to be disclosed after the vulnerability was introduced in a Rust package (median: 693, min: 2, max: 2,364, std: 534.2). As the per-vulnerability median indicates, 50\% of the vulnerabilities had disclosure duration exceeding 693 days (1.9 years). Our observations concur with prior findings that vulnerabilities in the npm ecosystem are disclosed within a median of 24 months, considerably shorter than the 37 months required for the vulnerabilities in the PyPI ecosystem~\cite{alfadel2021empirical}. 
The vulnerability with the longest disclosure duration (2,364 days) is RUSTSEC-2022-0029 in the \emph{crossbeam} package, introduced in December 2015 and disclosed in June 2022.
1.8\% of the vulnerabilities had a disclosure duration lower bound of less than 30 days, most of which are introduced after August 2021, indicating an increase in the security awareness within the Rust ecosystem.

In \autoref{fig:vul_fix_duration}, we depict the cumulative distribution function (CDF) of the number of days between disclosure and fixing.
The predominant behavior in \autoref{fig:vul_fix_duration}, observed for two-thirds of the vulnerabilities (120 out of 180), is that the vulnerability fixes were committed by disclosure time, manifesting as negative or zero time differences. The predominant behavior suggests that the majority of vulnerabilities in the Rust ecosystem were either internally discovered or disclosed to project developers using private channels, which is the expected best practice for vulnerability disclosure~\cite{zhou2021finding}.
In \autoref{fig:vul_fix_duration}, vulnerabilities disclosed but not yet fixed manifest as positive time difference values, which occurred for one-third of vulnerabilities (60 out of 180) in the Rust ecosystem. The percentage of unpatched vulnerabilities by disclosure is higher than the 21.2\% as reported in prior study~\cite{CCS2017Patches} and comparable to the 30\% for Windows vulnerabilities~\cite{frei2011end}.
The 60 vulnerabilities with positive time difference values have an average of 88 days of their fixing duration (median: 23.5, min: 1, max: 686, std: 146.0). The vulnerability with the longest fix duration (686 days) was RUSTSEC-2017-0006 in the \emph{rmpv} package, which was disclosed in November 2017 and fixed in October 2019.
In addition, approximately 42\% of the vulnerabilities remain unpatched for more than 30 days after their disclosure, leaving a window of opportunity for potential attacker exploitation. 

We further compared the duration of vulnerability disclosure and fixing across vulnerability types as shown in ~\autoref{tab:cve_type}. We observe that the duration of vulnerability disclosure and fixing vary widely across vulnerability types. Generally, frequently occurred vulnerability types (Count > 20) tend to have a significantly longer duration of disclosure compared to rarely occurred vulnerability types (Count < 20), as supported by Wilcoxon rank-sum tests (p-value = 0.0266). Among the rarely occurred vulnerability types, the \emph{API} vulnerabilities turn out to be an exception with the longest duration of disclosure and fixing across vulnerability types. 

\noindent\textbf{Vulnerability evolution.}
\begin{figure}[t]
\centering
    \begin{minipage}{0.48\textwidth}
        \centering
        \includegraphics[width=6.5cm]{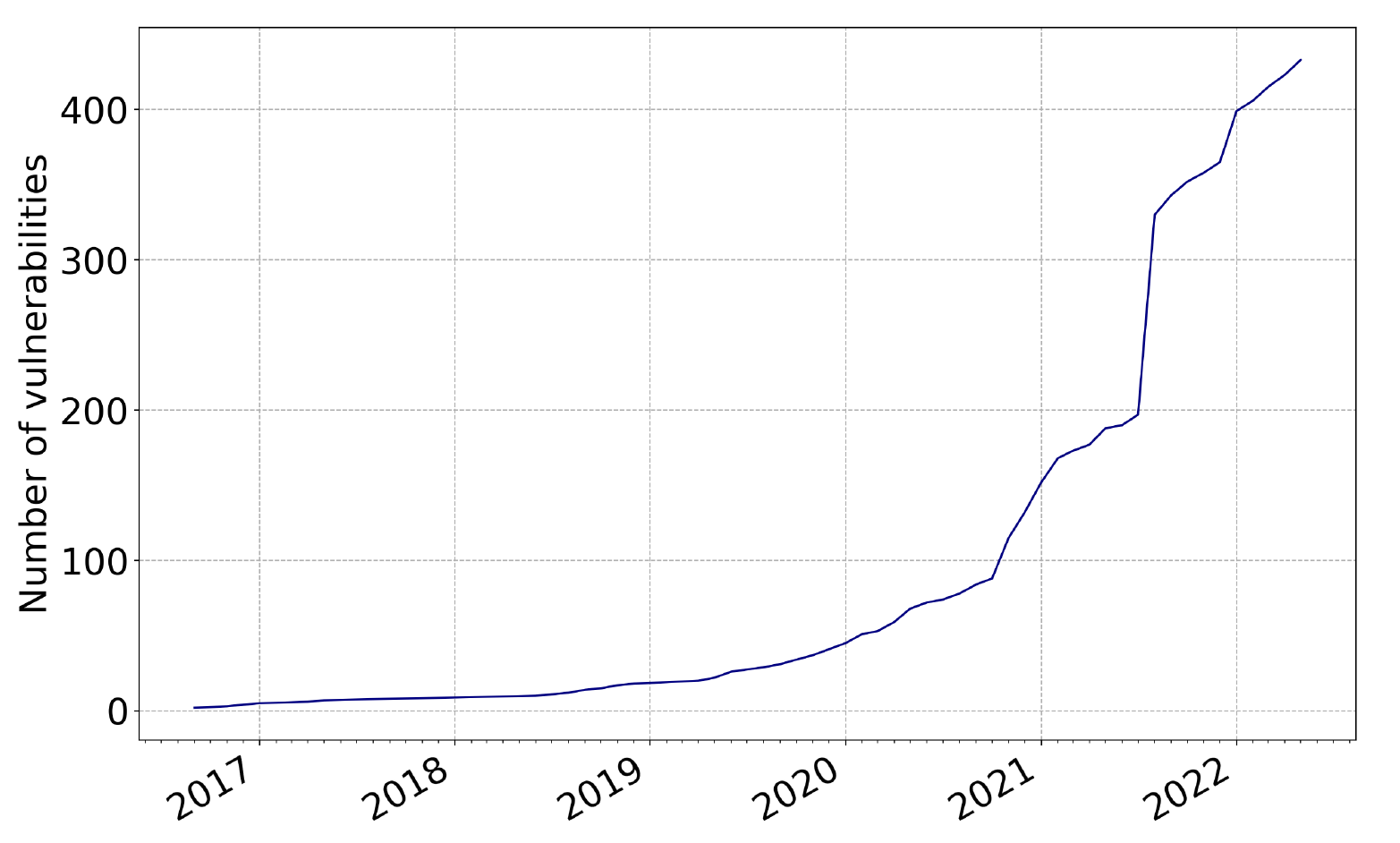}
\caption{Evolution of number of vulnerabilities disclosed over time.}
        \label{fig:numVulnerability}
    \end{minipage}
    \begin{minipage}{0.48\textwidth}
        \centering
        \includegraphics[width=\linewidth]{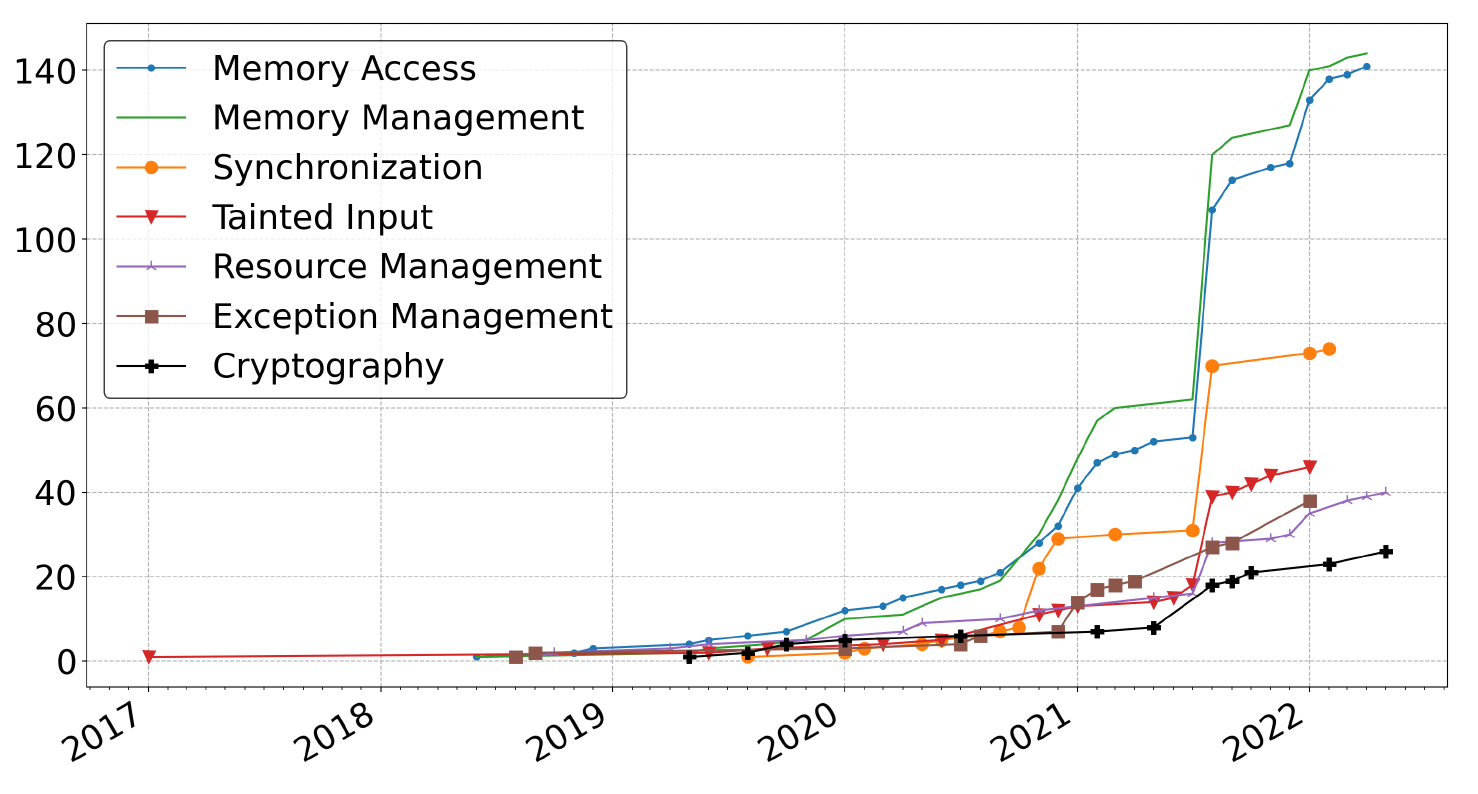}
        \caption{Evolution of numbers of disclosed vulnerabilities across vulnerability types.}
        \label{fig:numVulAcrossType}
    \end{minipage}
\end{figure}
We present the evolution of the number of vulnerabilities disclosed over time in the Rust ecosystem in \autoref{fig:numVulnerability}. We observe that the number of vulnerabilities disclosed grows slowly from November 2014 to November 2020, and experiences two rapid growth phases. The first rapid growth starts from November 2020 and ends in March 2021, while the second occurs in July 2021. The first rapid growth may attribute to a large-scale campaign during that period of time, in which RustSec has published 129 memory safety vulnerabilities as part of the research efforts made by Bae et al. ~\cite{Rudra}. During the second rapid growth, the number of disclosed vulnerabilities increased from 197 to 330 due to unrestricted {\tt Send} or {\tt Sync} on generic types, which is also discussed in prior work~\cite{Rudra}. 
The evolution in the numbers of vulnerabilities vary widely across vulnerabilities types as shown in \autoref{fig:numVulAcrossType}. The \emph{memory access}, \emph{memory management} and \emph{synchronization} vulnerabilities grow fastest over time, with a growth pattern that resembles the evolution of vulnerabilities disclosed in the Rust ecosystem as shown in \autoref{fig:numVulnerability}.

\begin{figure}[t]
\centering
    \begin{minipage}{0.48\textwidth}
        \centering
        \includegraphics[width=6.5cm]{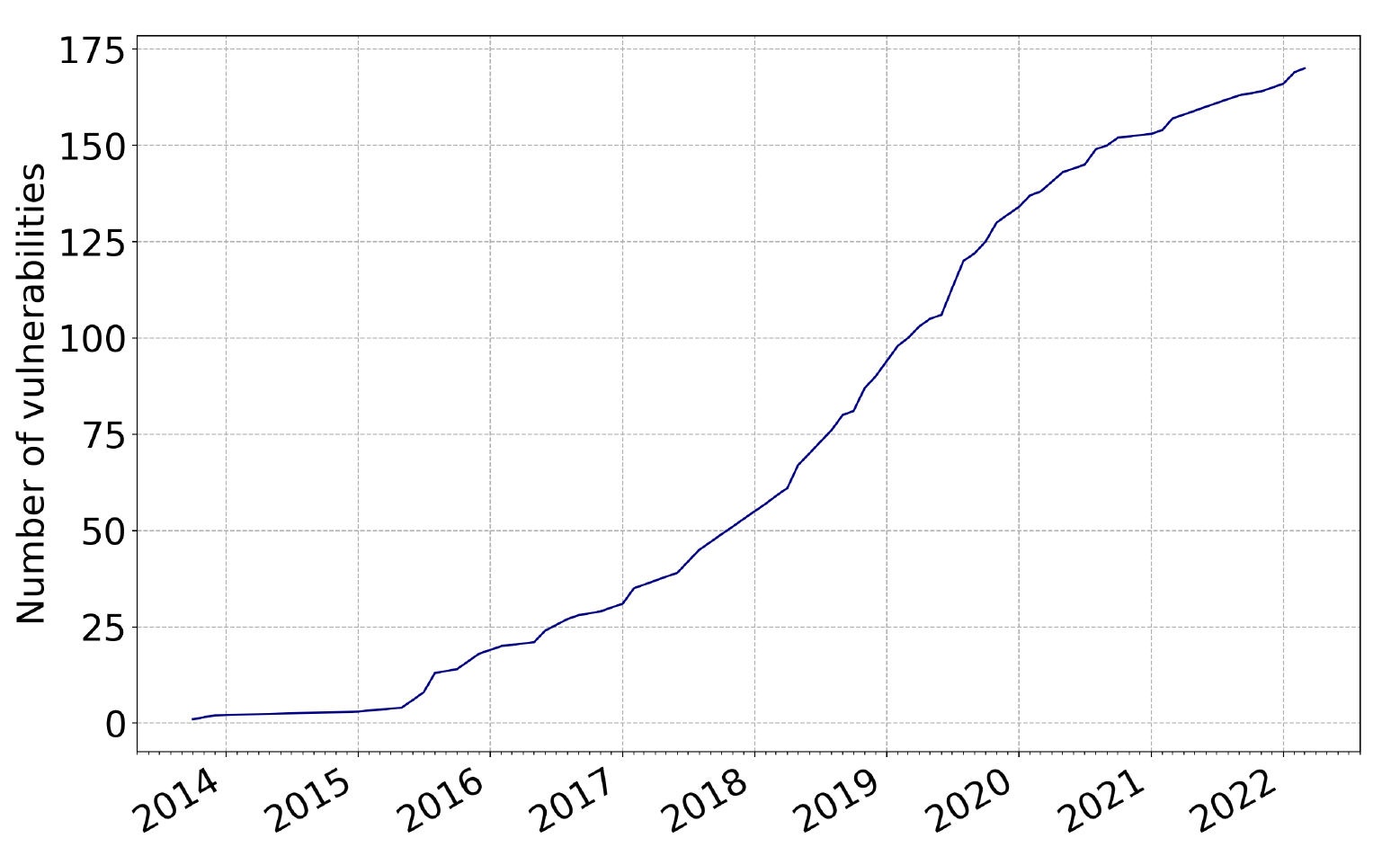}
        \caption{Evolution of number of vulnerabilities introduced into Rust code repositories over time.}
        \label{fig:introVulnerability}
    \end{minipage}
\end{figure}

We also present the evolution of the number of vulnerabilities introduced into Rust code repositories over time in \autoref{fig:introVulnerability}, which demonstrates a linear growth rate from July 2015 to January 2020 ($R^2$ = 0.9561) and becomes stabilized after March 2020. The numbers of vulnerabilities disclosed and introduced demonstrate different growth rates over time, which may be due to the increase of individuals and organizations participated in Rust vulnerability discovery as Rust becomes increasingly popular in systems software development. Another possible reason could be the development and application of vulnerability detection tools in the Rust ecosystem, which facilitate the discovery of vulnerabilities~\cite{Rudra,li2021mirchecker}.

\begin{figure}[t]
\centering
    \begin{minipage}{0.48\textwidth}
        \centering
        \includegraphics[width=6.2cm]{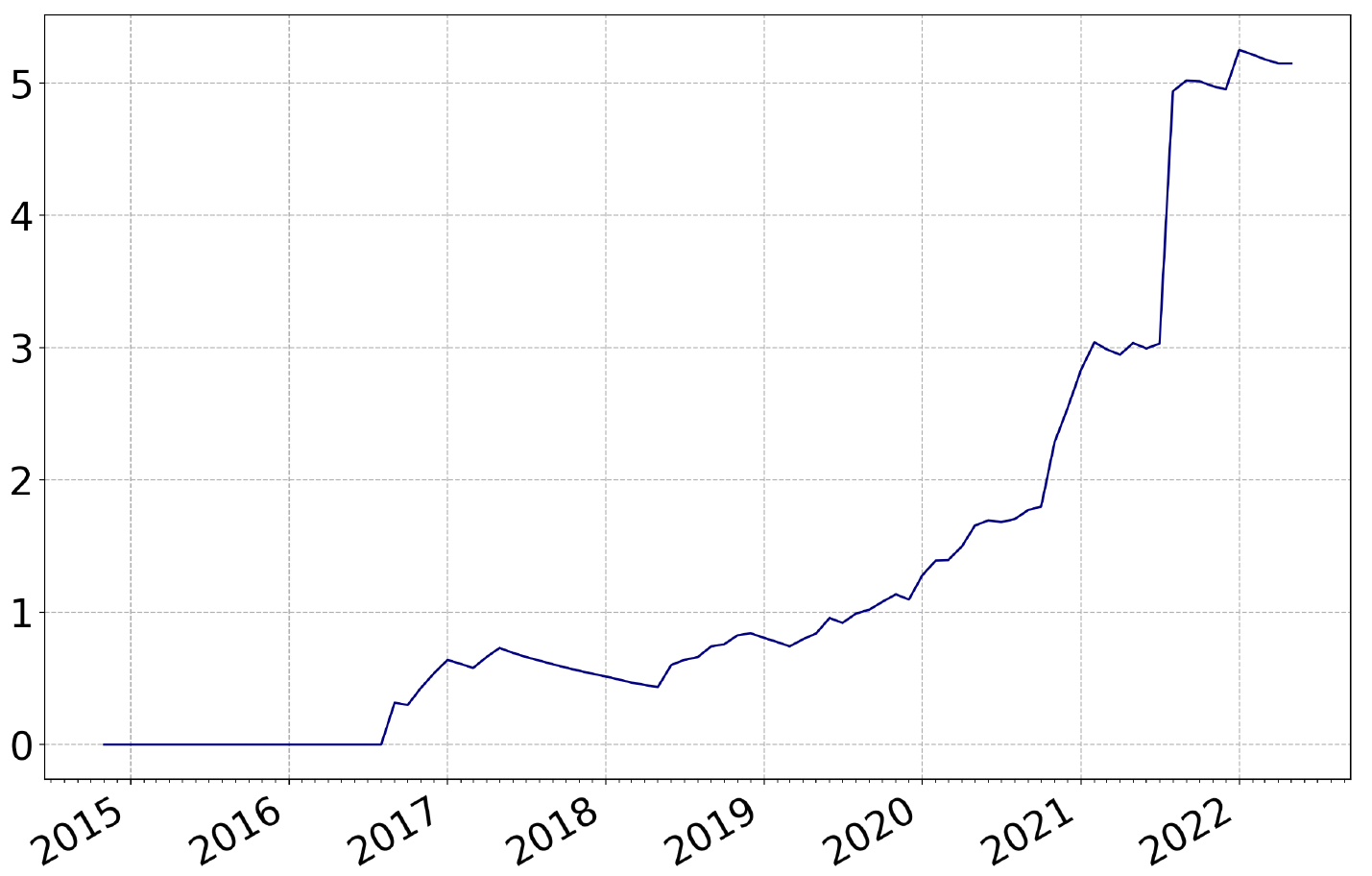}
        \caption{Disclosed vulnerabilities per 1,000 packages.}
        \label{fig:Vulnerabilityrate}
    \end{minipage}
    \begin{minipage}{0.48\textwidth}
    \vspace{10pt}
        \centering
        \includegraphics[width=6.5cm]{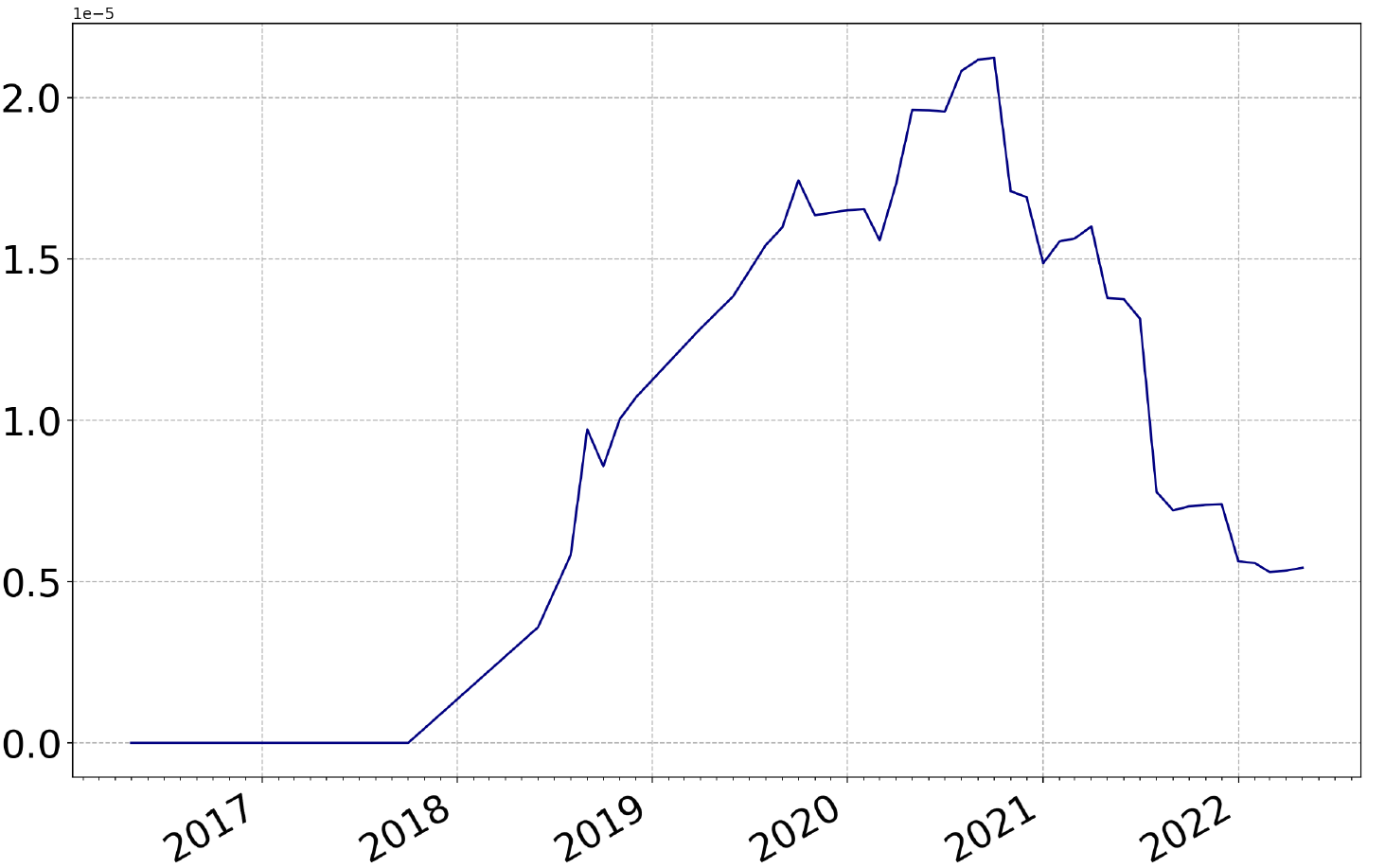}
        \caption{Disclosed vulnerabilities per 100,000 lines of code in vulnerable packages.}
        \label{fig:VulnerabilityLoc}
    \end{minipage}
\end{figure}

Next, we investigated whether package growth in the Rust ecosystem contributes to the increase of disclosed vulnerabilities. As shown in \autoref{fig:Vulnerabilityrate}, the normalized number of vulnerabilities disclosed per 1,000 packages grows from one in 2017 to five in 2022, indicating an increase of package-wise security risks in the Rust ecosystem. Meanwhile, the normalized number of vulnerabilities disclosed per 100,000 lines of code in vulnerable packages reaches the peak (2.1) in August 2020 after three climbing stages, and have experienced a sharp decrease to 0.5 since then, as shown in \autoref{fig:VulnerabilityLoc}, suggesting a decreasing tendency in security risks per lines of code in the Rust ecosystem.

\begin{figure}[t]
    \centering
    \captionsetup[subfigure]{font=tiny}
    \begin{subfigure}[b]{0.4\textwidth}
        \includegraphics[width=\linewidth]{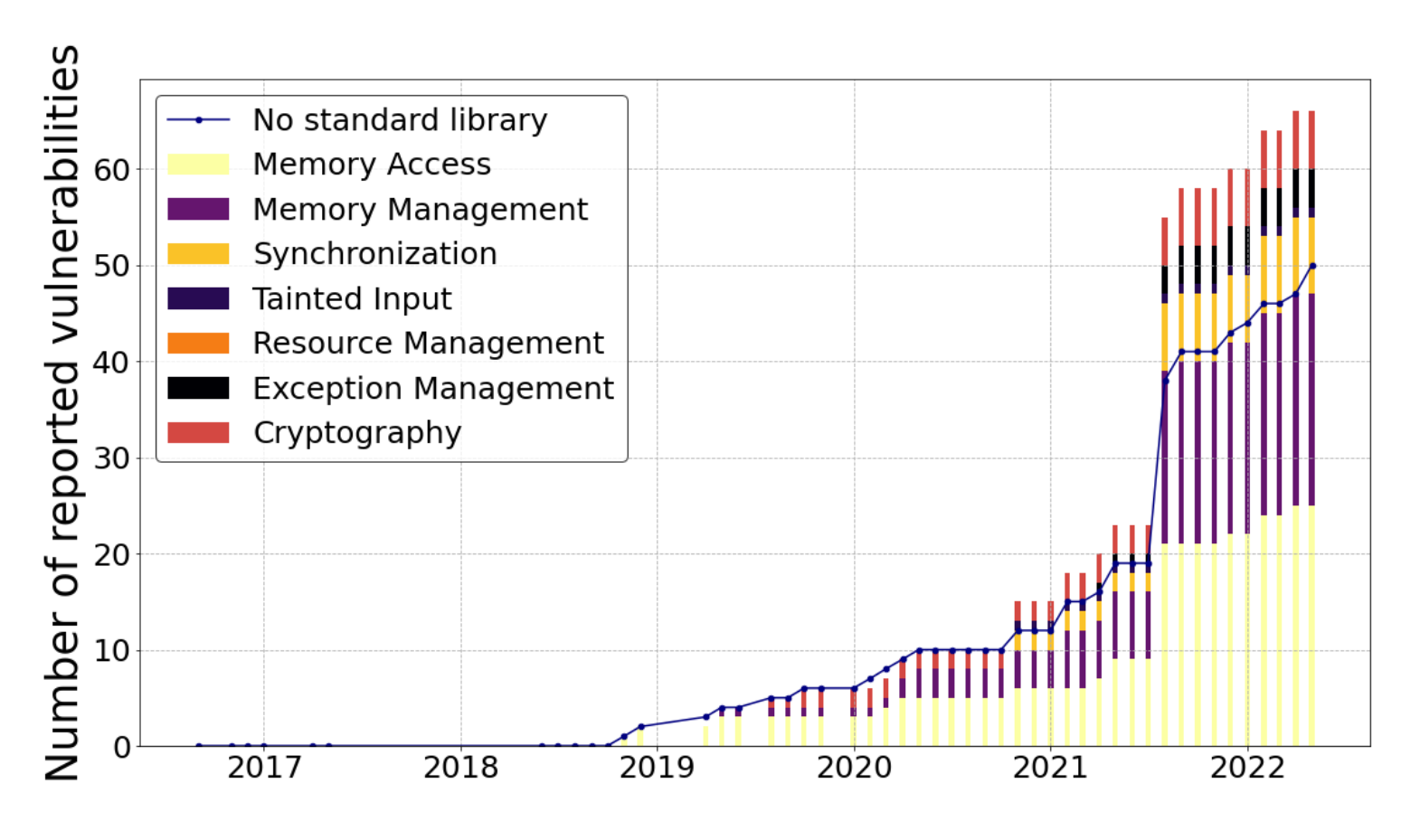}
        \caption{No standard library}
        \label{subfig:no-std}
    \end{subfigure}
\hfil
    \begin{subfigure}[b]{0.4\textwidth}
        \includegraphics[width=\linewidth]{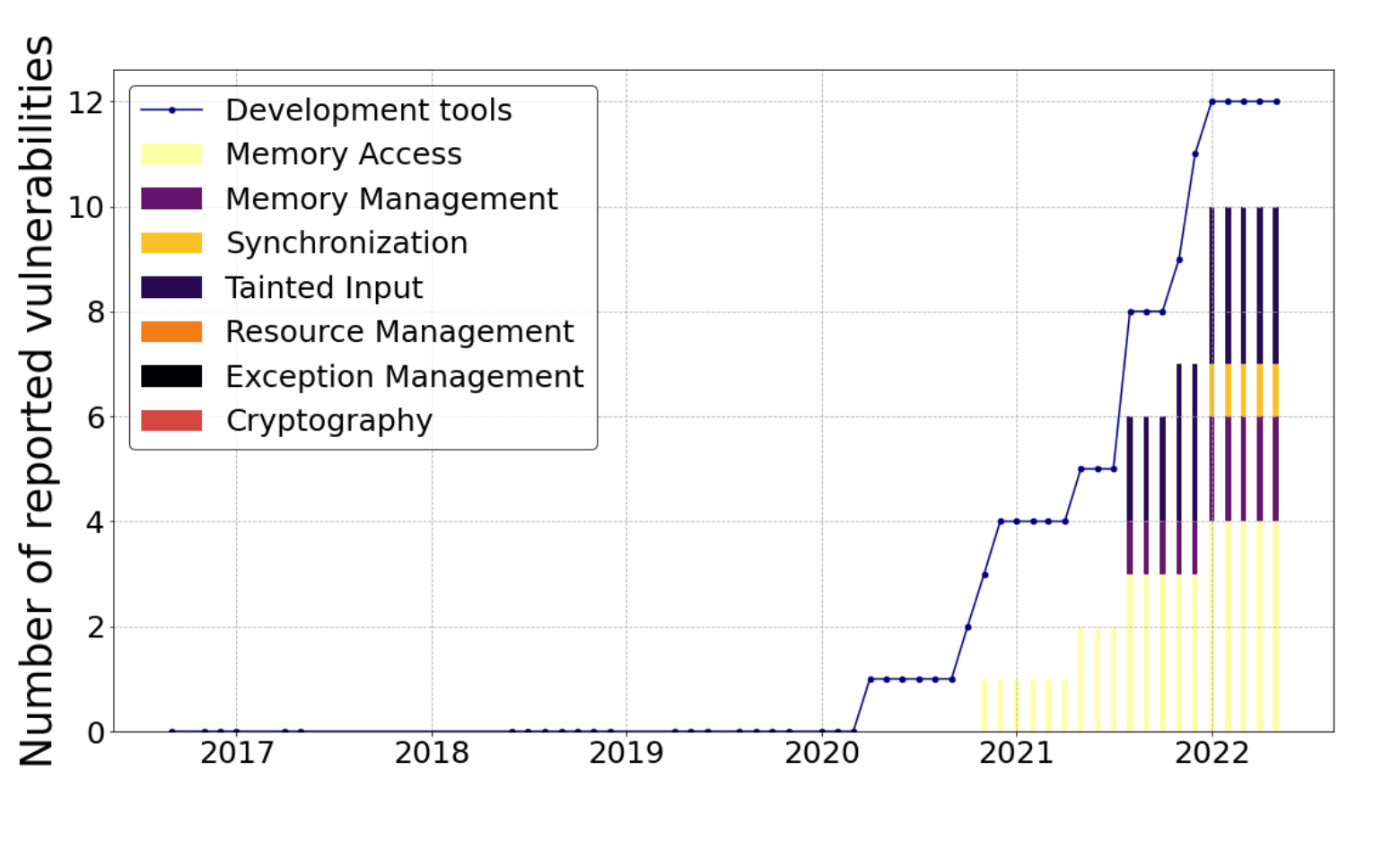}
        \caption{Development tools}
    \end{subfigure}

    \begin{subfigure}[b]{0.4\textwidth}
        \includegraphics[width=\linewidth]{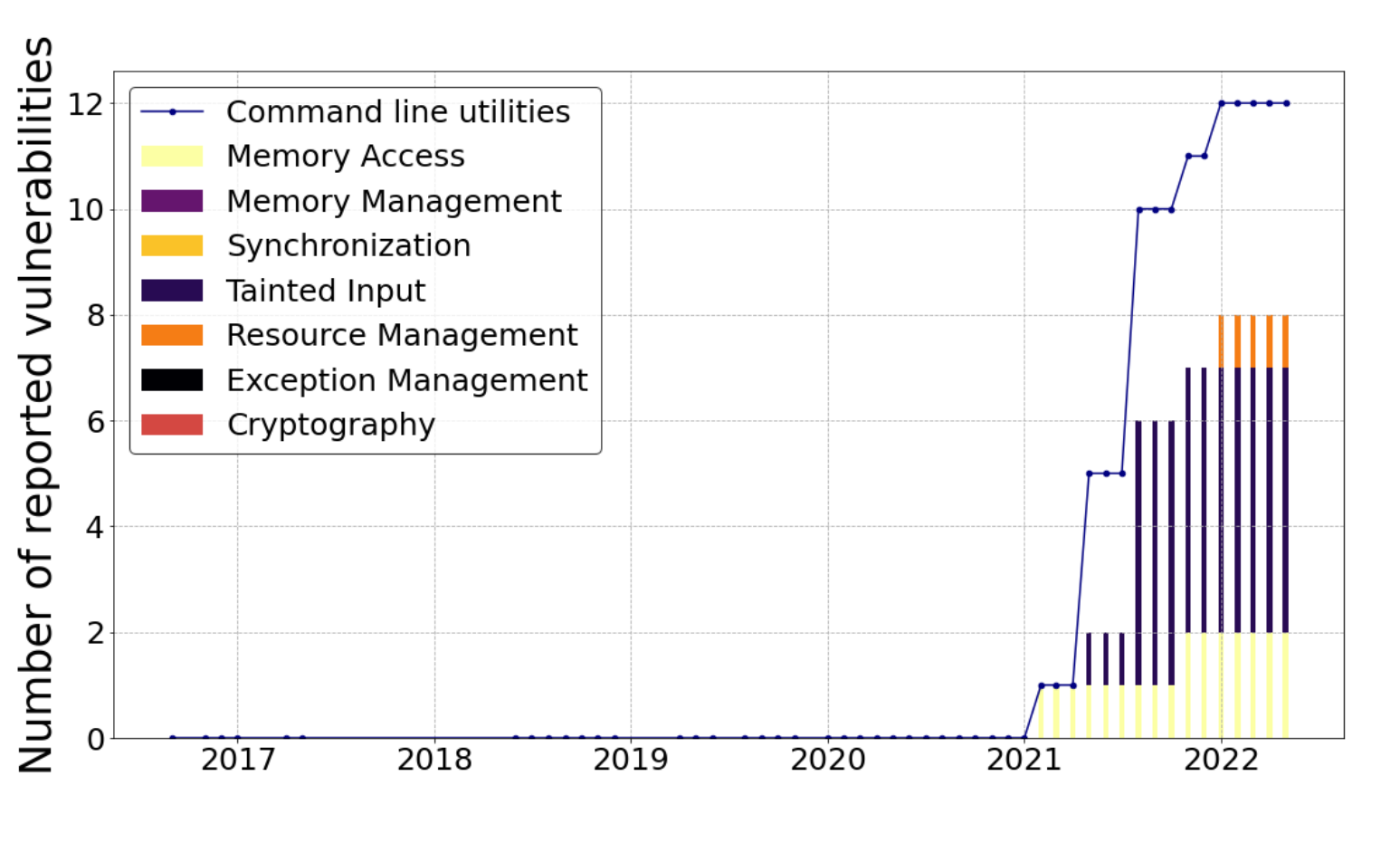}
        \caption{Command line utilities}
        \label{subfig:cmd}
    \end{subfigure}
\hfil
    \begin{subfigure}[b]{0.4\textwidth}
        \includegraphics[width=\linewidth]{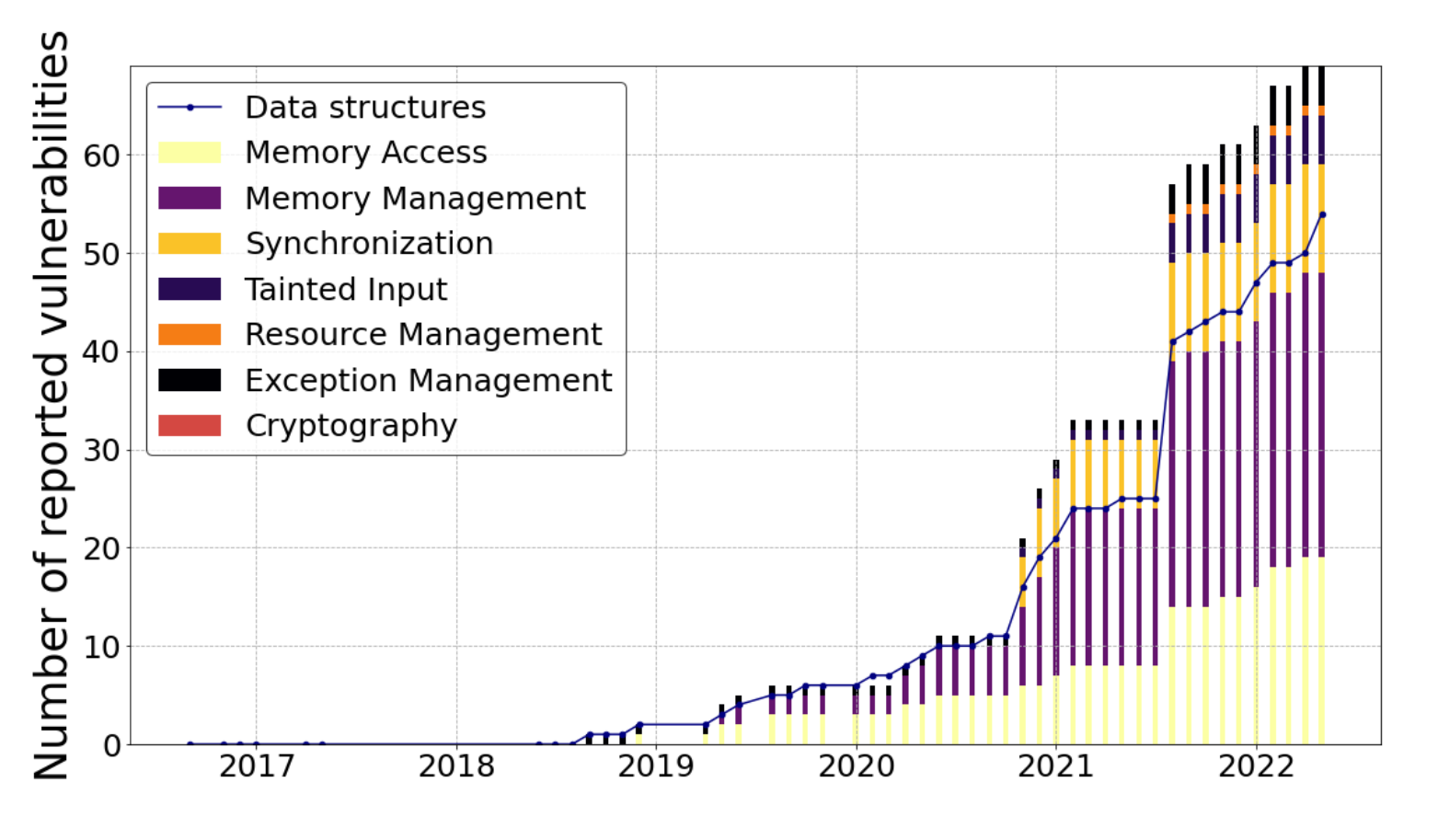}
        \caption{Data structures}
        \label{subfig:ds}
    \end{subfigure}

    \begin{subfigure}[b]{0.4\textwidth}
        \includegraphics[width=\linewidth]{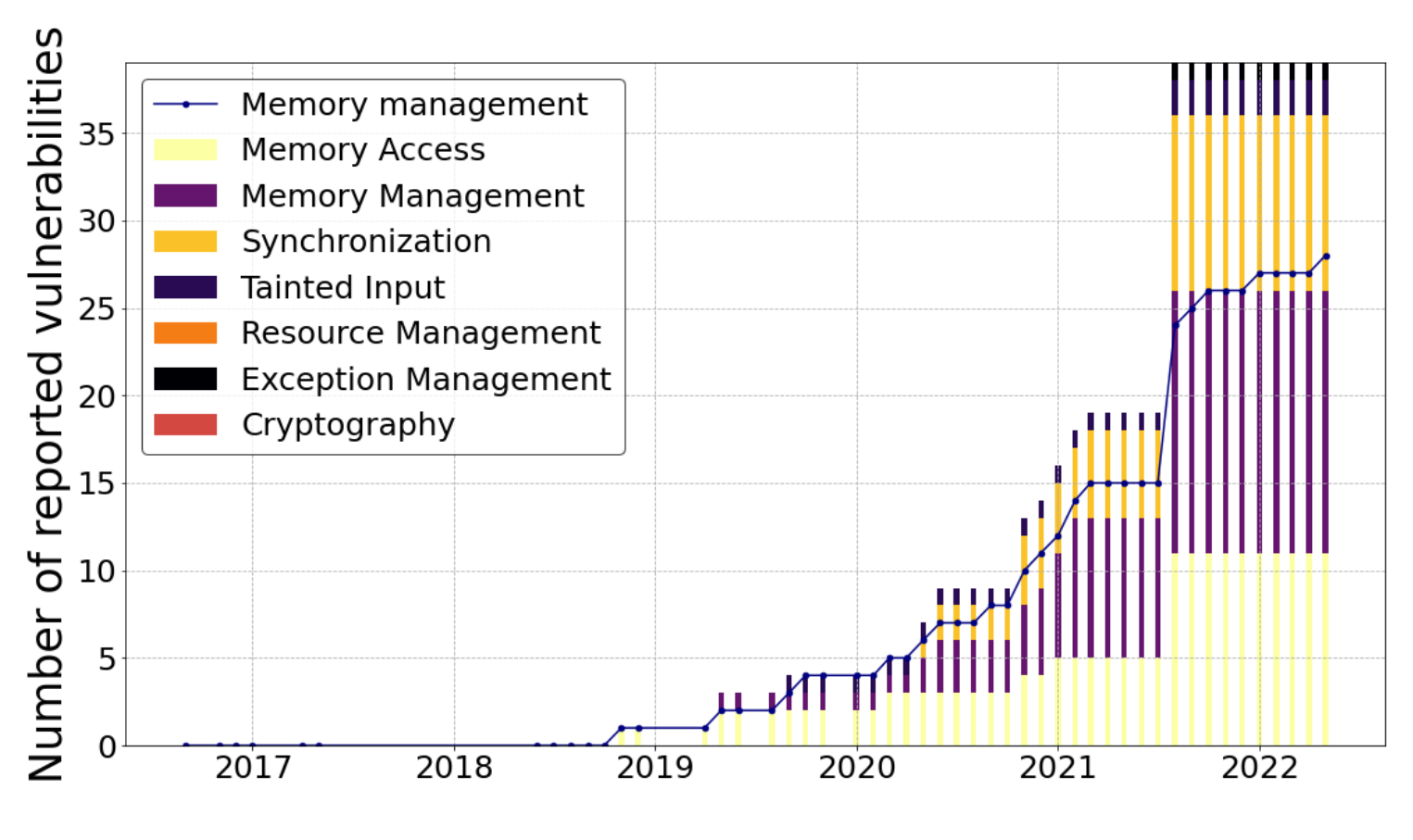}
        \caption{Memory management}
    \end{subfigure}
\hfil
    \begin{subfigure}[b]{0.4\textwidth}
        \includegraphics[width=\linewidth]{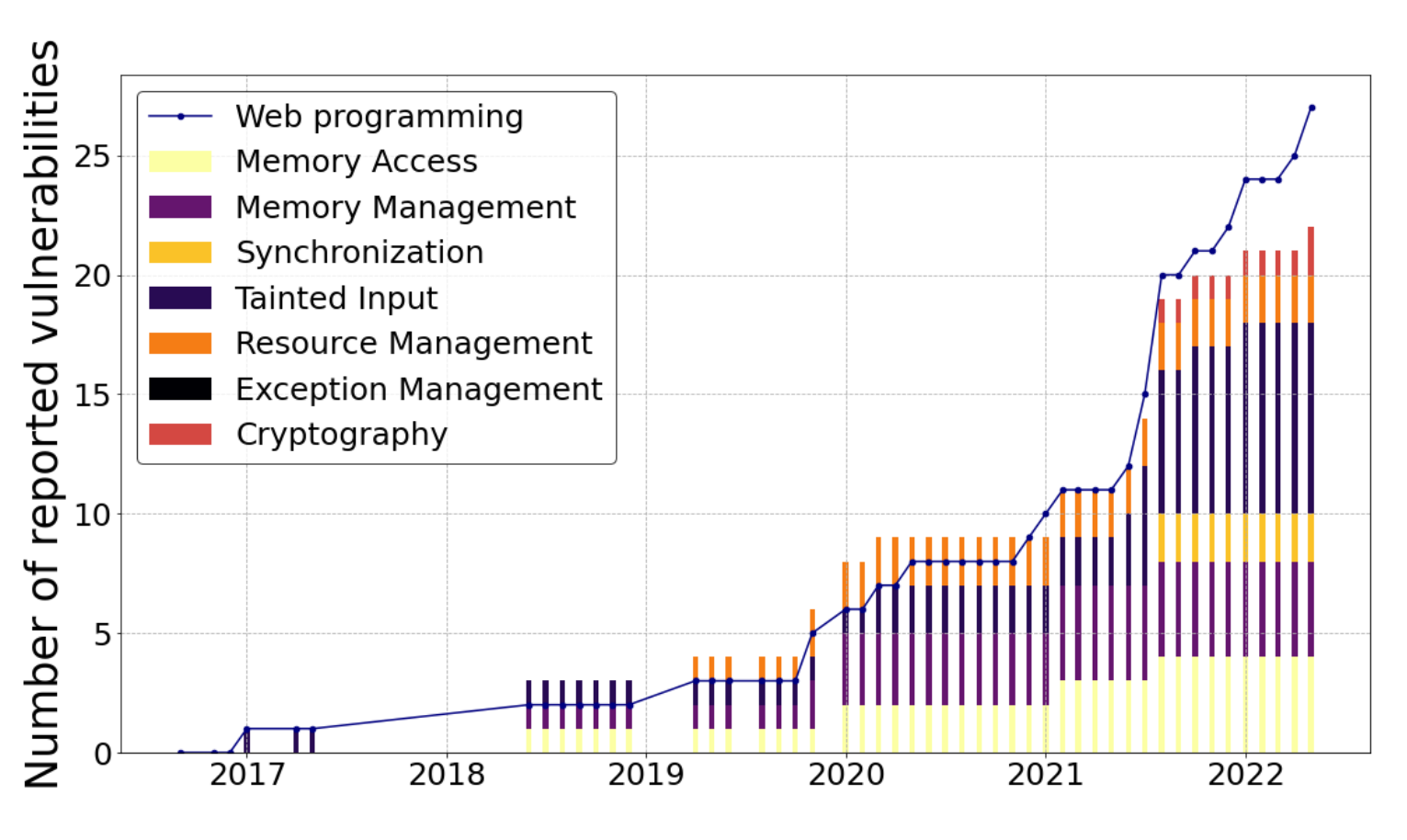}
        \caption{Web programming}
    \end{subfigure}

    \begin{subfigure}[b]{0.4\textwidth}
        \includegraphics[width=\linewidth]{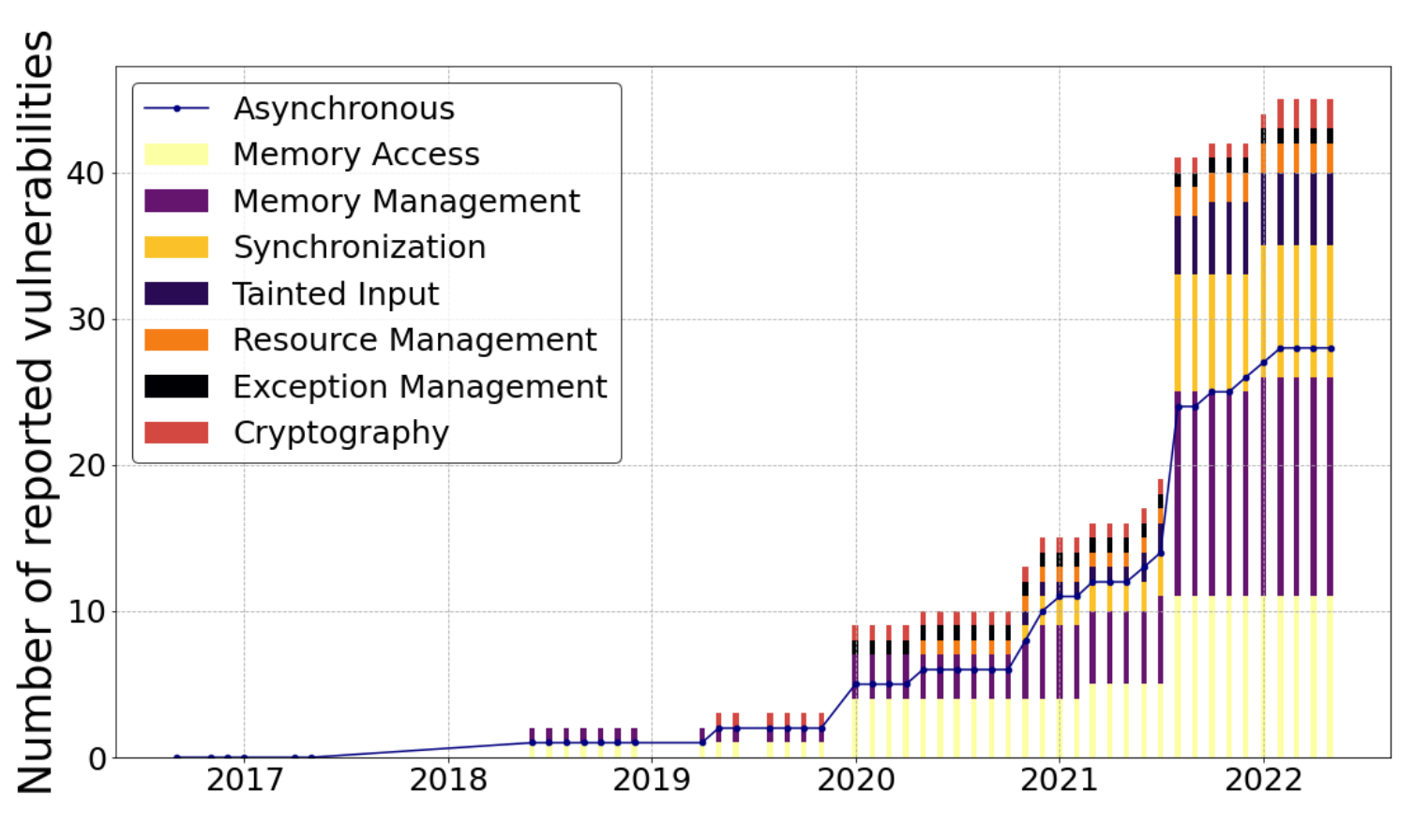}
        \caption{Asynchronous}
    \end{subfigure}
\hfil
    \begin{subfigure}[b]{0.4\textwidth}
        \includegraphics[width=\linewidth]{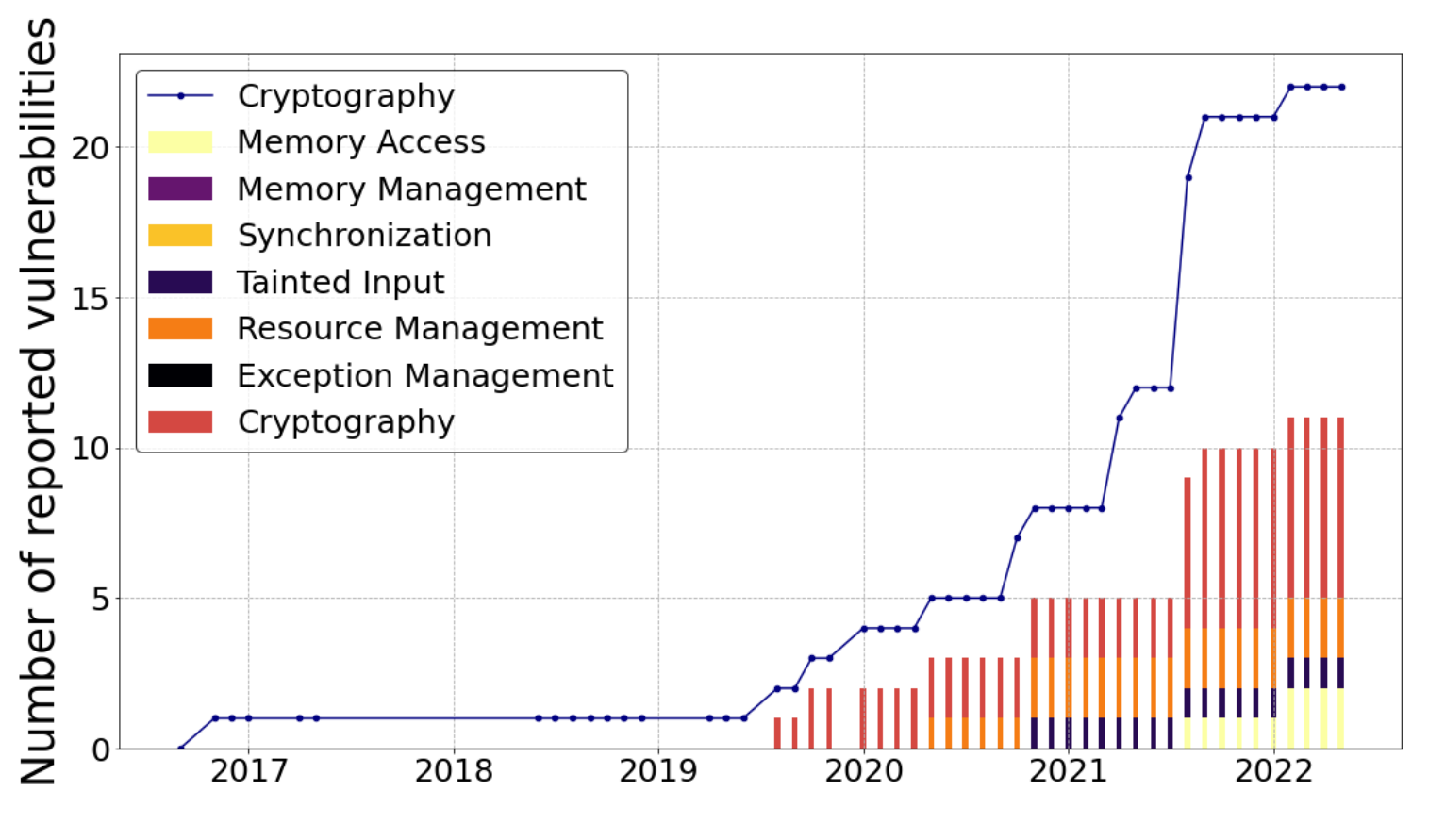}
        \caption{Cryptography}
    \end{subfigure}
    \caption{Evolution of numbers of disclosed vulnerabilities across package categories.}
    \label{fig:numVulAcrossCategory}
\end{figure}

Finally, we investigated how vulnerabilities with different types evolve over time with respect to package category as shown in \autoref{fig:numVulAcrossCategory}. We make several observations:
\begin{itemize}
\item \emph{Memory management} vulnerabilities are disclosed across package categories, but with different frequencies of occurrence and growth rates.
\item \emph{Command line utilities} packages have no vulnerability disclosed until April 2021, and have 12 vulnerabilities disclosed in total until April 2022, indicating relatively low security risk over time (\autoref{subfig:cmd}). 
\item \emph{No standard library} and \emph{data structures} packages both have more than 60 vulnerabilities disclosed in total until April 2022, among which \emph{memory management} vulnerabilities account for a majority (44.0\% in \emph{no standard library} and 53.7\% in \emph{data structures}) (\autoref{subfig:no-std} and \autoref{subfig:ds}). The numbers of disclosed vulnerabilities are greater than any other package categories, indicating relatively higher  security risk over time.
\end{itemize}

\vspace{-0.2cm}
\begin{tcolorbox}[width=\linewidth,
                  boxrule=0.5pt,
                  arc=0mm,
                  left=0pt,
                  right=0pt,
                  top=2pt,
                  colback=yellow!30]

\textbf{Summary for RQ1:} 
The top three vulnerability types in the Rust ecosystem are \emph{memory access}, \emph{memory management}, and \emph{synchronization}, accounting for 63.6\% of categorized vulnerabilities and exhibiting the fastest growth rates.
It takes over 2 years for the vulnerabilities to be publicly disclosed, among which 66.7\% have fixes committed before their disclosure. 
The number of disclosed vulnerabilities experiences two rapid growth periods, while the number of vulnerabilities introduced into code repositories grows linearly. 
Normalized numbers of disclosed vulnerabilities suggest a continuously increasing trend in package-level security risks over time, yet a decreasing trend in code-level security risks since August 2020. In addition, the security risks in the Rust ecosystem vary widely across different package categories. \end{tcolorbox}

\subsection{RQ2: Vulnerable Packages in the Rust Ecosystem}\label{sec:result_RQ2}
We identified a total of 337 vulnerable packages, accounting for 0.40\% of packages in the Rust ecosystem. 120 out of 337 vulnerable packages remain unpatched. The 337 vulnerable packages have an average of 1.3 disclosed vulnerabilities (min: 1, max: 14, median: 1, std: 1.04). The disclosed vulnerabilities affect an average of 28.6 versions of the Rust packages (min: 1, max: 339, median: 17, std: 34.33), accounting for 75.09\% of the versions per package on average (min: 1.37\%, max: 100\%, median: 82.07\%, std: 0.27). 

\noindent\textbf{Popularity of vulnerable packages.} On the one hand, popular packages tend to have more vulnerabilities. Specifically, the top 5 Rust packages with the most vulnerabilities are (1) {\tt openssl-src}, with 14 vulnerabilities and 55 versions affected (87.27\%), (2) {\tt wasmtime}, with 7 vulnerabilities and 62 versions affected (58.06\%), (3) {\tt hyper}, with 7 vulnerabilities and 224 versions affected (92.41\%), (4) {\tt ckb}, with 7 vulnerabilities and 32 versions affected (28.13\%), and (5) {\tt smallvec}, with 5 vulnerabilities and 54 versions affected (68.52\%). 
Among the top 5 packages, {\tt openssl-src}, {\tt hyper} and {\tt smallvec} have over 9 millions of downloads till April 2023. One possible reason could be that popular packages have far more developers and users than less popular ones. The larger community of developers and users for popular Rust packages tend to uncover more vulnerabilities compared to less popular packages. 
On the other hand, some popular packages officially published by September 2016 have no vulnerabilities disclosed, e.g., {\tt libc} and {\tt syn}, which have over 100 million downloads. 
In addition, we observe that unpopular packages (with less than 100,000 downloads) tend to suffer from vulnerabilities for more versions. Particularly, among the 109 vulnerable packages with all versions affected by vulnerabilities, 76 packages have less than 100,000 downloads in total over time.

\begin{table}[t]
\caption{Characteristics of package categories with vulnerable packages.}
\label{tab:category_percent}
\footnotesize
  \begin{tabular}{lcrrC{2.5cm}}
    \toprule
    \textbf{Package Category}    & \textbf{\# Vulnerabilities} &  \textbf{\# Packages} & \textbf{\# Downloads} & \textbf{\#  Vulnerabilities per Package}  \\
    \midrule
    Memory management       & 28    & 468       & 397,961,797       &5.98\%      \\
Concurrency             & 26    & 909       & 799,571,861       &2.86\%      \\
    Data structures         & 54    & 2,144     & 1,170,205,196     &2.52\%     \\
    Caching                 & 5    & 234     & 68,948,180           &2.14\%      \\
    Network programming     & 33    & 2,138     & 816,412,017       &1.54\%     \\
    Asynchronous            & 28    & 1,900     & 1,039,591,569     &1.47\%     \\
    Encoding                & 22    & 1,494     & 904,623,837       &1.47\%     \\
    No standard library     & 50    & 3,671     & 3,704,866,624     &1.36\%     \\
    Rust patterns           & 15    & 1,158     & 829,299,334       &1.30\%     \\
    Parsing tools           & 8     & 1,272     & 363,898,651      &1.26\%     \\
    Text processing         & 11     & 943       & 484,155,024      &1.17\%    \\
    Cryptography            & 22    & 1,933     & 912,432,279       & 1.14\%   \\
    Web programming         & 27    & 2,386     & 650,382,813       & 1.13\%   \\
    Algorithms              & 17    & 1,595     & 1,190,075,755     &1.07\%     \\
Operating systems       & 8     & 1,013       & 603,471,887    &0.79\%     \\
API bindings            & 10     & 2,265     & 417,282,873      &0.44\%    \\
    Development tools       & 12    & 3,725     & 1,640,269,851     &0.32\%     \\
    Command line utilities  & 12    & 4,418     & 25,115,904        &0.27\%     \\
    (vulnerabilities<=6)   & 90 & 14,670   & 2,461,533,553 &0.61\%     \\
    \emph{Non-categorized}      & 214 & 55,895   & -              &0.38\%     \\
    \bottomrule
  \end{tabular}
\end{table}

\noindent\textbf{Categories of vulnerable packages.}
\autoref{tab:category_percent} reports the numbers of disclosed vulnerabilities, packages, and downloads till May 2022, as well as the average numbers of vulnerabilities per package across package categories. The top 3 package categories with the most vulnerabilities are \textit{data structures}, \textit{no standard library} and \textit{network programming}. In the meantime, \textit{memory management}, \textit{concurrency} and \textit{data structures} rank the top 3 among package categories in terms of the average number of vulnerabilities per package. 
Interestingly, the \textit{memory management} category has fewer packages (468) and downloads (around 400 million downloads), but relatively more vulnerabilities disclosed (28), compared with other package categories, indicating that \textit{memory management} packages are more prone to vulnerabilities.

We further compare the distributions of vulnerability types across package categories. As shown in \autoref{tab:cve_tpe_across_type}, we chose the union of the top 5 categories with vulnerable package percentage and the top 5 categories with vulnerabilities in \autoref{tab:category_percent} for analysis. 
The distributions of vulnerability types vary substantially across package categories: 
The \emph{memory management} package category tends to have more \emph{memory access}, \emph{memory management} and \emph{synchronization} vulnerabilities; the \emph{concurrency} packages tend to have more \emph{synchronization} and \emph{memory management} vulnerabilities; and the \emph{data structure} packages tend to have more \emph{tainted input} vulnerabilities.
To see if the differences in the distributions of vulnerability types are statistically significant,  we conduct Wilcoxon signed-rank tests~\cite{wilcoxon} with Bonferroni correction
at 95\% significance level. As a result, we observe statistically significant differences (1) \emph{caching} vs. \emph{data structures} (p-value = 0.0490) and (2) \emph{caching} vs. \emph{no standard library} (p-value = 0.0490), indicating the penitential impact of package categories on disclosed vulnerability types in the Rust ecosystem.

\begin{table}[t]
  \caption{Distributions of vulnerability types in top 5 package categories with vulnerabilities and top 5 package categories with vulnerable package percentages.}
  \label{tab:cve_tpe_across_type}
  \footnotesize
\begin{tabulary}{\textwidth}{lp{1.2cm}p{1.2cm}p{1.2cm}p{1cm}p{1.2cm}p{1.2cm}}
    \toprule
     & \textbf{\scriptsize Memory Management} & \textbf{\scriptsize Concurrency} & \textbf{\scriptsize Data Structures} & \textbf{\scriptsize Caching} & \textbf{\scriptsize No Standard Library}&\textbf{\scriptsize Network Programming}\\
    \textbf{Vulnerability Type} & & & & & & \\
    \midrule
    Memory Access           & 11       & 8 & 19 & 2 & 25 & 11\\
    Memory Management       & 15       & 17 & 29 & 3 & 22 & 10\\
    Synchronization         & 10       & 17 & 11 & 4 & 8 & 4\\
    Tainted Input           & 2        & 4  & 5 & 0 & 1 & 3\\
    Resource Management     & 0        & 1  & 1 & 0 & 0 & 4\\
    Exception Management    & 1        & 0  & 4 & 0 & 4 & 3\\
    Cryptography            & 0        & 0  & 0 & 0 & 6 & 3\\
    Other                   & 0        & 0  & 3 & 0 & 4 & 2\\
    Risky Values            & 2        & 0  & 4 & 0 & 3 & 1\\
    Path Resolution         & 1        & 0  & 1 & 0 & 0 & 3\\
Predictability          & 0         & 0  & 0 & 0 & 2 & 0\\
API                     & 0         & 0  & 0 & 0 & 1 & 0\\
Failure to Release Memory & 0       & 0  & 1  & 0 & 0  & 0\\
    \emph{Total}                & 28    & 26 & 54 & 5 & 50 & 33\\
    \bottomrule
  \end{tabulary}
\end{table}

\begin{table}[t]
  \caption{Descriptive statistics of vulnerability locality in vulnerable packages.}
  \label{tab:vul_loc}
  \footnotesize
  \begin{tabular}{lcccc}
    \toprule
 & \textbf{\# Files} & \textbf{\# Safe Functions} & \textbf{\# Unsafe Functions} & \textbf{\# Unsafe Blocks} \\
    \midrule
    mean ($\mu$)    & 1.85  & 3.35  & 0.15  & 1.39\\
    median (M)      & 1     & 1     & 0     & 0\\
    min             & 1     & 0     & 0     & 0\\
    max             & 14    & 83    & 4    & 50\\
    std             & 1.95  & 8.84  & 0.53  & 5.20\\
    total           & 395   & 684   & 31   & 284\\
    \bottomrule
  \end{tabular}
\end{table}

\begin{figure}[t]
\centering
    \begin{minipage}{0.48\textwidth}
        \centering
        \includegraphics[width=6.5cm]{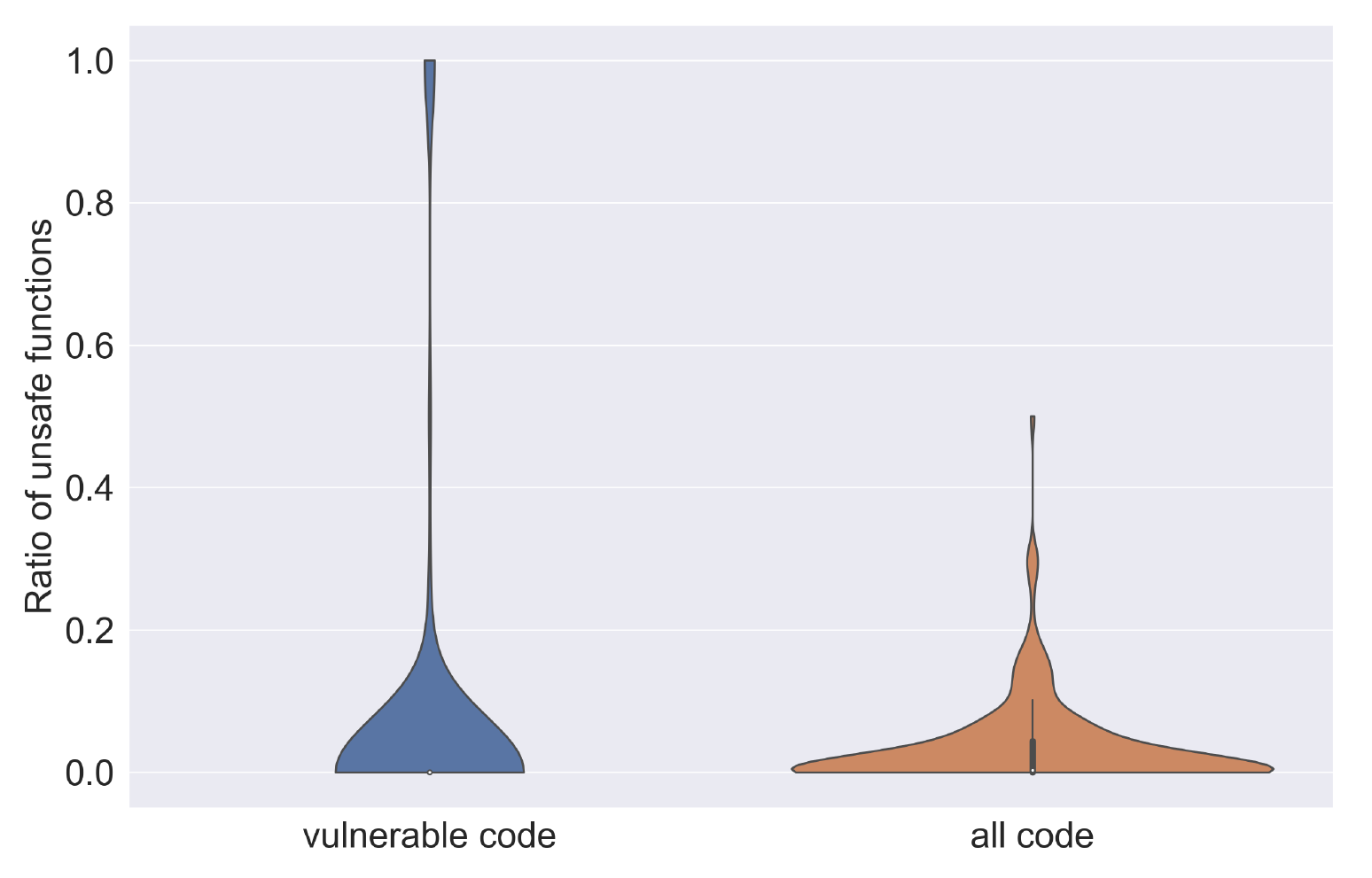}
        \caption{Ratios of unsafe functions in vulnerable code vs. all code of vulnerable packages.}
\label{fig:disFunc}
    \end{minipage}
    \begin{minipage}{0.48\textwidth}
        \centering
        \includegraphics[width=7cm]{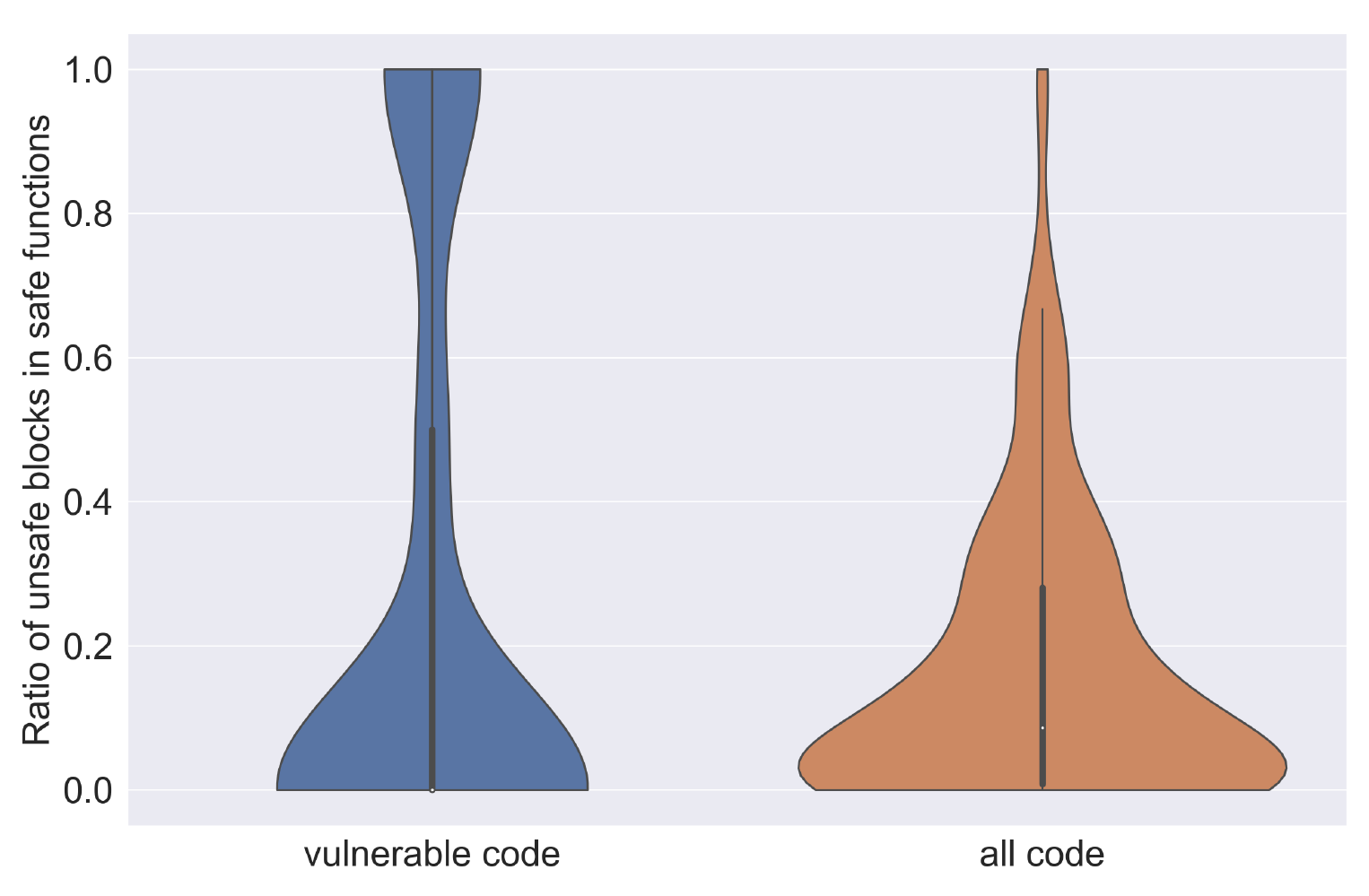}
        \caption{Ratios of unsafe blocks in vulnerable code vs. all code of vulnerable packages.}
\label{fig:disBlock}
    \end{minipage}
\end{figure}

\noindent\textbf{Vulnerability locality in vulnerable packages.} 
As illustrated in \autoref{tab:vul_loc}, a disclosed vulnerability affects 1.85 files, 3.35 safe functions, 0.15 unsafe functions, and 1.39 unsafe blocks on average in vulnerable packages in the Rust ecosystem (1 file, 1 safe function, 0 unsafe function, and 0 unsafe block in median). The small median number of affected files per fix commit indicates that vulnerable code in the Rust ecosystem is localized at the file level. At functional level, 95\% of the affected functions are safe functions; Among the affected safe functions, 41.5\% contain unsafe blocks in their body of a function. One possible reason for the high percentage of safe functions in vulnerable code is that developers tend to wrap unsafe code in safe functions and provide conditional checks in those functions before entering the unsafe code, which is in line with the Rust idiomatic style to encapsulate unsafety~\cite{Rudra}.

We further compared the ratios of unsafe functions and unsafe blocks between vulnerable code vs. all code in vulnerable packages as shown in \autoref{fig:disFunc} and \autoref{fig:disBlock}, respectively. 
We observe that vulnerable packages have higher ratios of unsafe functions (0.059 vs. 0.034 on average) and unsafe blocks (0.261 vs. 0.165 on average) in their vulnerable code as compared to their complete code. 
The Wilcoxon signed-rank test at 95\% significance level suggests statistically significant differences exist in the ratios of unsafe functions (p-value = 0.002) and unsafe blocks (p-value = 0.017) between vulnerable code and complete code in vulnerable packages.

Finally, we compared the vulnerability localities across vulnerability types, including the numbers of commits, files, safe and unsafe functions, and unsafe blocks, as shown in \autoref{tab:fun_cnt_across_type}. We make the following observations:
\begin{itemize}
\item The \emph{exception management} vulnerabilities tend to be the least localized at the file and function levels, considering the greatest numbers of files and safe functions affected by them as compared to other vulnerability types. 
\item The \emph{synchronization} vulnerabilities tend to be the most localized at the function level, considering the smallest number of safe functions on average they affected. \item The \emph{memory management} and \emph{memory access} vulnerabilities show the tendency to locate more frequently in safe functions than in unsafe functions, considering they affected more safe functions than unsafe ones. 
\item The \emph{resource management} and \emph{path resolution} vulnerabilities originate from the safe code in safe functions, considering they affected neither unsafe functions nor unsafe blocks.
\end{itemize}

\begin{table}[t]
\caption{Descriptive statistics of vulnerability locality across vulnerability types.}
  \label{tab:fun_cnt_across_type}
\tiny
\begin{tabular}{lllllllll}
\toprule
\multicolumn{2}{l}{\textbf{Vulnerability type}} &
  \textbf{\begin{tabular}[c]{@{}l@{}}Memory \\ Access\end{tabular}} &
  \textbf{\begin{tabular}[c]{@{}l@{}}Memory \\ Management\end{tabular}} &
  \textbf{Synchronization} &
  \textbf{\begin{tabular}[c]{@{}l@{}}Tainted \\ Input\end{tabular}} &
  \textbf{\begin{tabular}[c]{@{}l@{}}Resource \\ Management\end{tabular}} &
  \textbf{\begin{tabular}[c]{@{}l@{}}Exception \\ Management\end{tabular}} &
  \textbf{\begin{tabular}[c]{@{}l@{}}Path \\ Resolution\end{tabular}}\\\midrule
\multicolumn{2}{l}{\textbf{\# Commits}}                                   & 42    & 28   & 39   & 21    & 9   & 6   & 6\\ \midrule
\multicolumn{1}{l}{\multirow{5}{*}{\textbf{\# Files}}}            & $\mu$ & 2.30  & 2.14 & 1.59 & 1.57  & 1.89& 2.33& 1.83 \\ \multicolumn{1}{l}{}                                              & M     & 1     & 1    & 1    & 1     & 2    & 2  & 1.5 \\ \multicolumn{1}{l}{}                                              & min   & 1     & 1    & 1    & 1     & 1    & 1  & 1 \\ \multicolumn{1}{l}{}                                              & max   & 14    & 14   & 5    & 4     & 3    & 5  & 3\\ \multicolumn{1}{l}{}                                              & std   & 2.94  & 2.72 & 1.25 & 0.95  & 0.93 & 1.51 & 0.98\\ \midrule
\multicolumn{1}{l}{\multirow{5}{*}{\textbf{\# Safe functions}}}   & $\mu$ & 3.88  & 4.32 & 1.05 & 1.43  & 2 & 8.67 & 1.33\\ \multicolumn{1}{l}{}                                              & M     & 1     & 1    & 0    & 1     & 2    & 3    & 1\\ \multicolumn{1}{l}{}                                              & min   & 0     & 0    & 0    & 0     & 1    & 1    & 1\\ \multicolumn{1}{l}{}                                              & max   & 59   & 33   & 13   & 5     & 5   & 40    & 3\\ \multicolumn{1}{l}{}                                              & std   & 10.92 & 8.56 & 2.67 & 1.12  & 1.32 & 15.41 & 0.82\\ \midrule
\multicolumn{1}{l}{\multirow{5}{*}{\textbf{\# Unsafe functions}}} & $\mu$ & 0.17  & 0.14 & 0.15 & 0.05  & 0   & 0 & 0\\ \multicolumn{1}{l}{}                                              & M     & 0     & 0    & 0    & 0     & 0    & 0   & 0 \\ \multicolumn{1}{l}{}                                              & min   & 0     & 0    & 0    & 0     & 0    & 0   & 0 \\ \multicolumn{1}{l}{}                                              & max   & 2    & 3    & 2   & 1     & 0    & 0   & 0 \\ \multicolumn{1}{l}{}                                              & std   & 0.44  & 0.59 & 0.49 & 0.22  & 0    & 0 & 0\\ \midrule
\multicolumn{1}{l}{\multirow{5}{*}{\textbf{\# Unsafe blocks}}}    & $\mu$ & 2.14  & 2.39 & 0.21 & 0.19  & 0 & 7.67 & 0\\ \multicolumn{1}{l}{}                                              & M     & 0     & 0    & 0    & 0     & 0    & 2  & 0  \\ \multicolumn{1}{l}{}                                              & min   & 0     & 0    & 0    & 0     & 0    & 0   & 0 \\ \multicolumn{1}{l}{}                                              & max   & 50    & 67    & 6   & 3     & 0    & 40   & 0 \\ \multicolumn{1}{l}{}                                              & std   & 7.85  & 6.32 & 0.98 & 0.68  & 0 & 15.87 & 0\\
\bottomrule
\end{tabular}
\end{table}

\begin{tcolorbox}[width=\linewidth,
                  boxrule=0.5pt,
                  arc=0mm,
                  left=0pt,
                  right=0pt,
                  top=2pt,
                  colback=yellow!30]

\textbf{Summary for RQ2:} 
Vulnerable packages in the Rust ecosystem have an average of 1.3 disclosed vulnerabilities that affected 28.6 versions of the corresponding packages. Popular packages tend to have more vulnerabilities, while less popular ones tend to suffer from vulnerabilities for more versions. 
\emph{Memory management} is the most vulnerable package category, with a small number of packages but a large number of vulnerabilities. In addition, the \emph{memory management} package category tends to have more \emph{memory access}, \emph{memory management}, and \emph{synchronization} vulnerabilities as compared to other package categories. The \emph{exception management} vulnerabilities tend to be the least localized at the file and function levels as compared to other vulnerability types.
In vulnerable packages, the vulnerable code tends to involve statistically significantly more unsafe functions and unsafe blocks as compared to complete code, which is localized at the file level.
\end{tcolorbox}
\par

\begin{table}[t]
\caption{Descriptive statistics of vulnerability fixes.}
  \label{tab:commit_loc}
\footnotesize
\begin{tabular}{lcc|ccc}
    \toprule
& \textbf{\# Lines Added} & \textbf{\# Lines Deleted} & \textbf{\# Safe Functions} & \textbf{\# Unsafe Functions} & \textbf{\# Unsafe Blocks} \\
    \midrule
    mean ($\mu$)    & 41.13     & 18.17      & 3.85  & 0.16  & 1.49\\
    median (M)      & 14        & 4          & 1     & 0     & 0\\
    min             & 1         & 0          & 0     & 0     & 0\\
    max             & 665       & 330        & 83    & 4    & 50\\
    std             & 81.97     & 39.75      & 9.12  & 0.53  & 5.24\\
    total           & -         & -          & 786   & 32   & 304\\
    \bottomrule
  \end{tabular}
\end{table}

\subsection{RQ3: Vulnerability Fixes in the Rust Ecosystem}
\noindent\textbf{Fix commit complexity and locality.}
As shown in \autoref{tab:commit_loc}, the commits of vulnerability fixes in the Rust ecosystem involve an average of 41 and 18 LOC added and deleted, respectively (14 and 4 LOC added and deleted in median). The commits of vulnerability fixes have touched 3.85 safe functions, 0.16 unsafe functions, and 1.49 unsafe blocks on average. Safe functions account for 96\% of all the functions touched by the vulnerability fixes, among which 38.8\% contain unsafe blocks.

Next, we compared the locality of vulnerability fix commits across vulnerability types with respect to the numbers of safe functions, unsafe functions, and unsafe blocks touched by the commits as shown in \autoref{tab:commit_fun_cnt_across_type}. The fix commits of \emph{exception management} vulnerabilities involve the most safe functions compared to other vulnerability types (8.67), indicating the widest spread of touched code and potential challenges when fixing \emph{exception management} vulnerabilities in practice. 
On the contrary, the fix commits of \emph{synchronization} vulnerabilities touch the fewest safe functions compared to other vulnerability types (1.21), indicating the most localized fixes across vulnerability types.

\begin{table}[t]
\caption{Statistics of commit locality of vulnerability fixes across vulnerability types.} \label{tab:commit_fun_cnt_across_type}
  \tiny
\begin{tabular}{lllllllll}
\toprule
\multicolumn{2}{l}{\textbf{Vulnerability Type}} &
  \textbf{\begin{tabular}[c]{@{}l@{}}Memory \\ Access\end{tabular}} &
  \textbf{\begin{tabular}[c]{@{}l@{}}Memory \\ Management\end{tabular}} &
  \textbf{Synchronization} &
  \textbf{\begin{tabular}[c]{@{}l@{}}Tainted \\ Input\end{tabular}} &
  \textbf{\begin{tabular}[c]{@{}l@{}}Resource \\ Management\end{tabular}} &
  \textbf{\begin{tabular}[c]{@{}l@{}}Exception \\ Management\end{tabular}} &
  \textbf{\begin{tabular}[c]{@{}l@{}}Path \\ Resolution\end{tabular}}\\\midrule
\multicolumn{1}{l}{\multirow{5}{*}{\textbf{\# Safe Functions}}}   & $\mu$ & 4.0   & 4.71 & 1.21 & 2.0  & 3.44 & 8.67    &2.17\\ \multicolumn{1}{l}{}                                              & M     & 1     & 2    & 0    & 1    & 2    & 3       & 2 \\ \multicolumn{1}{l}{}                                              & min   & 0     & 0    & 0    & 0    & 1    & 1       &1 \\ \multicolumn{1}{l}{}                                              & max   & 59    & 33   & 16   & 8    & 12   & 40      &5\\ \multicolumn{1}{l}{}                                              & std   & 10.89 & 8.60 & 3.22 & 1.79 & 3.50 & 15.41   & 1.47\\ \midrule
\multicolumn{1}{l}{\multirow{5}{*}{\textbf{\# Unsafe Functions}}} & $\mu$ & 0.19  & 0.14 & 0.15 & 0.05 & 0    & 0       & 0\\ \multicolumn{1}{l}{}                                              & M     & 0     & 0    & 0    & 0    & 0    & 0       & 0   \\ \multicolumn{1}{l}{}                                              & min   & 0     & 0    & 0    & 0    & 0    & 0       & 0   \\ \multicolumn{1}{l}{}                                              & max   & 2     & 3    & 2    & 1    & 0    & 0       & 0 \\ \multicolumn{1}{l}{}                                              & std   & 0.45  & 0.59 & 0.49 & 0.22 & 0    & 0       & 0\\ \midrule
\multicolumn{1}{l}{\multirow{5}{*}{\textbf{\# Unsafe Blocks}}}    & $\mu$ & 2.17  & 2.39 & 0.21 & 0.24 & 0    & 7.67    & 0 \\ \multicolumn{1}{l}{}                                              & M     & 0     & 0    & 0    & 0    & 0    & 2       & 0  \\ \multicolumn{1}{l}{}                                              & min   & 0     & 0    & 0    & 0    & 0    & 0       & 0  \\ \multicolumn{1}{l}{}                                              & max   & 50    & 33   & 6    & 3    & 0    & 40      & 0  \\ \multicolumn{1}{l}{}                                              & std   & 7.84  & 6.32 & 0.98 & 0.70 & 0    & 15.87   & 0\\ 
\bottomrule
\end{tabular}
\end{table}

\begin{table}[t]
  \centering
\caption{The characteristics of vulnerability fix commits across localities.}
\footnotesize
    \begin{tabular}{p{0.5cm}lr}
    \toprule
    \multicolumn{1}{c}{\textbf{Vulnerability Locality}} & \multicolumn{1}{l}{\textbf{Fix Pattern}} & \multicolumn{1}{c}{\textbf{Count}} \\
    \midrule
    \multicolumn{1}{l}{Safe function} & add safe functions & 28 \\
          & remove safe functions &  12 \\
          & modify safe functions & 117 \\
    \midrule
    \multicolumn{1}{l}{Unsafe function} & add unsafe functions &  0\\
          & remove unsafe functions & 0 \\
          & modify unsafe functions & 16 \\
    \midrule
    \multicolumn{1}{l}{Unsafe block} & add functions or lines in function & 5 \\
          & remove unsafe blocks &  27\\
          & modify unsafe blocks &  36\\
    \bottomrule
    \end{tabular}\label{tab:fix_pattern}
\end{table}

\noindent\textbf{Vulnerability fix patterns.}
To capture how developers fix different types of vulnerabilities, we first compared the numbers of safe functions, unsafe functions, and unsafe blocks affected by vulnerabilities (\autoref{tab:fun_cnt_across_type}) with those touched by corresponding fixes (\autoref{tab:commit_fun_cnt_across_type}). 
We make the following observations: (1) The \emph{tainted input} vulnerabilities demonstrate the greatest increase in the number of unsafe blocks compared to other vulnerability types, from 0.19 to 0.24; The increase indicates that developers tend to add unsafe blocks when fixing \emph{tainted input} vulnerabilities. (2) The \emph{path resolution} vulnerabilities have the greatest increase in the number of safe functions, from 1.33 to 2.17, indicating the addition of safe functions when fixing such vulnerabilities. (3) The \emph{path resolution} vulnerabilities have the greatest increase in the number of unsafe functions, from 0.17 to 0.19, indicating the addition of unsafe functions when fixing such vulnerabilities.

We further summarized the resulting fix patterns of our our manual inspection with respect to different vulnerability localities as shown in \autoref{tab:fix_pattern}. 
In general, we observe three operations, i.e., \emph{addition}, \emph{deletion}, and \emph{modification} of code, when developers fix vulnerabilities. Among the three operations, the modification operation accounts for the majority of fix commits across different vulnerability localities. We also make the following observations in particular:

\noindent\textbf{Observation 1: Developers tend to add safe functions, or add lines in safe functions to fix vulnerable safe functions.} Developers tend to add safe code when fixing vulnerabilities that locate in safe functions, rather than removing existing code. 

The added safe functions or lines validate pre-conditions or customize the default implemented functions, thus fix corresponding vulnerabilities.
Taking the fix of the \emph{denial-of-service} vulnerability (CVE-2022-24713)
as an example (\autoref{lst:add_func_example}), the fix inserts an additional function {\tt c\_empty()} to deal with an empty string, which may cause denial of service. 

\begin{minipage}{\linewidth}
\begin{lstlisting}[caption={Example vulnerability fix ( CVE-2022-24713).},label={lst:add_func_example},language=Rust,basicstyle=\small\ttfamily]
// code snapshot before fix commit
fn c(&mut self, expr: &Hir) -> ResultOrEmpty {
    ...
    match *expr.kind() {
        Empty => Ok(None),
        ...
    }
}

// code snapshot after fix commit
fn c_empty(&mut self) -> ResultOrEmpty {
    self.extra_inst_bytes += 
        std::mem::size_of::<Inst>();
    Ok(None)
}
fn c(&mut self, expr: &Hir) -> ResultOrEmpty {
    ...
    match *expr.kind() {
        Empty => self.c_empty(),
        ...
    }
}
\end{lstlisting}
\end{minipage}

\noindent\textbf{Observation 2: Developers tend to remove unsafe blocks to fix vulnerable unsafe blocks.}  Developers tend to remove existing code, rather than adding code to fix vulnerabilities that locate in unsafe blocks. Taking the fix of the memory exposure vulnerability (RUSTSEC-2021-0086) as an example (\autoref{lst:remove_unsafe}), the fix removes the unsafe block and zeroes out the buffer {\tt buf} before further operations can be undertaken.

\begin{minipage}{\linewidth}
\begin{lstlisting}[caption={Example vulnerability fix ( RUSTSEC-2021-0086).},label={lst:remove_unsafe},language=Rust,basicstyle=\small\ttfamily]
// code snapshot before fix commit
let mut buf = Vec::with_capacity(frame.data_size);
unsafe { buf.set_len(frame.data_size) };
    
// code snapshot after fix commit
let mut buf = vec![0; frame.data_size];
\end{lstlisting}
\end{minipage}

\noindent\textbf{Observation 3: Developers tend to modify unsafe trait implementations to fix vulnerable unsafe functions.} Specifically, we identified a \textit{data race} pattern that occurs in disclosed vulnerabilities that locate in unsafe functions as shown in \autoref{lst:send_sync}. In such vulnerability pattern, objects do not restrict to sendable or syncable types when they implement {\tt Send} or {\tt Sync} traits, leading to the sharing of non-syncable types across threads in concurrent programs. To fix such vulnerabilities, developers tend to use {\tt Send} or {\tt Sync} traits as bounds to stipulate the functionality that the generic type {\tt T} must implement.

\begin{minipage}{\linewidth}
\begin{lstlisting}[caption={Vulnerability fix pattern of \emph{data race} vulnerabilities.},label={lst:send_sync},language=Rust,basicstyle=\small\ttfamily]
// code snapshot before fix commit
unsafe impl<T> Send for MyObject<T> {}
unsafe impl<T> Sync for MyObject<T> {}
    
// code snapshot after fix commit
unsafe impl<T: Send> Send for MyObject<T> {}
unsafe impl<T: Sync> Sync for MyObject<T> {}
\end{lstlisting}
\end{minipage}

\begin{tcolorbox}[width=\linewidth,
                  boxrule=0.5pt,
                  arc=0mm,
                  left=0pt,
                  right=0pt,
                  top=2pt,
                  colback=yellow!30]

\textbf{Summary for RQ3:} 
The vulnerability fix commits in the Rust ecosystem involve a median of 14 lines of code added, and a median of 4 lines of code deleted, suggesting that vulnerability fix commits are typically localized.
96\% of the functions touched by these fix commits are safe functions.
The fix commits of \emph{exception management} vulnerabilities involve the highest proportion of safe functions compared to other vulnerability types, indicating potential challenges of fixing \emph{exception management} vulnerabilities in practice. 
In addition, developers tend to (1) add safe functions or add lines in safe functions to fix vulnerable safe functions, (2) remove unsafe blocks to fix vulnerable unsafe blocks, and (3) modify unsafe trait implementations to fix vulnerable unsafe functions.
\end{tcolorbox}

\section{Discussion}\label{sec:discussion}
We now summarize our main results, discuss their implications,  and highlight the avenues for future research.

\noindent\textbf{Rust is an active and growing software ecosystem at its early stage, coupled with an increasing awareness of risks of security vulnerabilities.}
The Rust ecosystem hosts over 100 thousand packages on {\tt crates.io} by March 2023, and has been experiencing exponential growth in number of packages and downloads from 2014 to 2022 as observed in our preliminary investigation (Section 2.3). As compared to other ecosystems like npm, PyPI and Maven, which host over 2  million packages\footnote{\url{http://www.modulecounts.com/}}, 440 thousand packages\footnote{\url{https://pypi.org/}}, and 32 million artifacts\footnote{\url{https://mvnrepository.com/repos}}, respectively by March 2023, we observed a significantly smaller number of packages in the Rust ecosystem, indicating Rust is at its early age of development. In addition, the exponential growth in number of packages in the Rust ecosystem resembles the trends of package growth at the early ages of the PyPI~\cite{Bommarito2018PyPI} and Maven ecosystems~\cite{mvn2022repo}, suggesting Rust to be an active and growing ecosystem.

Interestingly, our preliminary investigation observed a sharp decline in the increasing number of Rust packages since mid-late 2020, which may be caused by the Mozilla lay-off in (August 2020)\footnote{\url{https://blog.rust-lang.org/2020/08/18/laying-the-foundation-for-rusts-future.html}}.
The Rust packages in the \emph{no standard library} category, ranked the second in terms of the total number across package categories, received the most downloads over time. 
Future work could systematically investigate the factors that influence the evolution of the number of packages, and analyze how the sub-categories in the \emph{no standard library} category could affect the evolution of package downloads in the Rust ecosystem.

The number of vulnerabilities disclosed per 1,000 packages in the Rust ecosystem grows from one in 2017 to five in 2022 (RQ1), indicating an increasing awareness of the risks of security vulnerabilities in the ecosystem. The increasing trends in the number of disclosed vulnerabilities are observed in the PyPI and npm ecosystems as well~\cite{Bommarito2018PyPI}, which may attribute to the coordinated efforts in the increasing awareness of security risks in the ecosystems and continuous process of testing packages to detect vulnerabilities before exploited.

\noindent\textbf{The majority of the vulnerabilities in the Rust ecosystem relate to memory safety and concurrency issues.}
Memory safety and concurrency issues account for two-thirds of the vulnerabilities in the Rust ecosystem (RQ1).
The frequent occurrence of memory safety and concurrency issues may be due to that developers tend to use the Rust programming language for systems software development, thus manipulating memory and threads in their code frequently.
In contrast, cross-site scripting vulnerabilities appear to be the most common type of vulnerability in both PyPI and npm ecosystems as discussed in prior research~\cite{Bommarito2018PyPI}, given Python and JavaScript are popular programming languages for the development of Web applications.
We also find that 77\% of the vulnerabilities related to memory safety and concurrency issues locate in unsafe code of vulnerable Rust packages (RQ2), suggesting that practitioners should pay more attention to operations related to memory and concurrency when writing unsafe code. For instance, practitioners could enforce bound constraints when implementing {\tt Send} and {\tt Sync} traits to avoid data race as observed in RQ3.

\noindent\textbf{Vulnerabilities in the Rust ecosystem are not localized in unsafe functions, but relate more to unsafe blocks in safe functions as compared to the safe code in safe functions.}
Safe functions account for 95\% of the functions in the vulnerable packages affected by disclosed vulnerabilities (RQ2). The percentage of safe functions in the vulnerable packages affected by disclosed vulnerabilities is close to the percentage of safe functions in the Rust ecosystem as reported in a recent study (95.9\%)~\cite{unsafeblog}, indicating that vulnerabilities in the Rust ecosystem are not localized in unsafe functions.
Meanwhile, 41.5\% of the safe functions in the vulnerable packages affected by disclosed vulnerabilities contain unsafe blocks in their function body (RQ2). The frequency of occurrence of unsafe blocks in safe functions in the vulnerable packages is significantly higher than that of all the Rust packages (13.8\%) as reported in a recent study~\cite{unsafeblog}, indicating that practitioners could put forth more effort on unsafe blocks for securing safe functions compared to their safe code.

\noindent\textbf{{Practices towards safer Rust code.}}
86.67\% of the \emph{tainted input} and \emph{resource management} vulnerabilities reside in safe code of vulnerable packages (RQ2), indicating that the \emph{tainted input} and \emph{resource management} vulnerabilities are more localized in safe code than other types of vulnerabilities. The results suggest that practitioners should pay more attention to safe code for data validation when dealing with user input as compared to unsafe code.
In terms of fixing vulnerabilities, prior study~\cite{CCS2017Patches} reported that the median security commit diff involved 7 LOC. The results of RQ3 revealed that the median fix commit diff of the Rust ecosystem includes 14 and 4 LOC added and deleted, respectively, indicating that the fixes in Rust are more complex. Moreover, the study~\cite{CCS2017Patches} found that 59\% of security changes were located in a single function, while the results of RQ3 showed that the median fix commit involves an average of 3.85 safe functions (median = 3.85), suggesting that the fixes are less localized in the Rust ecosystem.
In addition, developers tend to modify both safe and unsafe code in vulnerable packages, rather than directly remove unsafe code when fixing vulnerabilities (RQ3), suggesting the necessity of unsafe code in the development of Rust packages, which is in line with the findings of a previous study~\cite{icse2020evan} -- practitioners use the unsafe keyword in 29.4\% of the Rust libraries. Thus, both safe and unsafe code deserve a comparable consideration when securing Rust packages in practice.

\section{Threats to Validity}\label{sec:threat}
\noindent\textbf{Construct validity.}
Our dataset contains the complete list of disclosed vulnerabilities in the Rust ecosystem by May 24, 2022, thus it is inevitable that the characteristics of vulnerabilities and fixes in the Rust ecosystem may evolve along with the expansion of vulnerability dataset over time. We analyzed vulnerability fix commits for the investigation of vulnerable packages and fixes in the Rust ecosystem, which is in line with previous studies~\cite{SZZ,tufano2018empirical,ray2016naturalness,chen2021neural}. The approach may introduce noise into the dataset given that some code inside the fix commits might not be vulnerable or fix vulnerabilities. To mitigate the threat, we manually inspected the 287 fix commits we have collected and excluded 69 fix commits with irrelevant code, e.g., refactoring and typo fixes. As a result, the vulnerability fix commits in our dataset involve a median of 14 and 4 LOC added and deleted, respectively, whose sizes are comparable to vulnerability fix commits in previous studies~\cite{CCS2017Patches, repeatedBugFix}. 

\noindent\textbf{Internal validity.}
We implemented a Rust compiler plugin and all scripts by ourselves with careful review; despite the extensive testing phase, we may not exclude all possible implementation errors. For the sake of replicability, we made all data and scripts employed publicly available\footnote{\url{https://github.com/ZXXYy/rust_ecosystem}}. Some steps in our methodology rely on information produced by the Rust compiler, e.g., identifying the localities of Rust vulnerabilities. Consequently, these steps may be sensitive to unfixed bugs in the Rust compiler.

For the 33 fix commits (15.14\%) that fail the compilation process, we used text analysis to collect the numbers of safe/unsafe functions and unsafe blocks in their code. To estimate the potential threat introduced by the text analysis, we used the 185 fix commits that pass the compilation process and compared their numbers of safe/unsafe functions and unsafe blocks identified by text analysis with those identified by the compiler plugin. We found that text analysis identifies 23.7\% less unsafe blocks, 10.5\% less safe functions, and 14.4\% less unsafe functions compared to the compiler plugin. As a result, text analysis would underestimate the total numbers of unsafe blocks, safe functions and unsafe functions by 3.6\%, 1.6\% and 2.2\% in our dataset, respectively. The ratio of unsafe block and unsafe function in all code of vulnerable package would reduce by 2\% and 0.6\%, respectively. Thus, the slight underestimation of safe/unsafe functions and unsafe blocks would not affect our conclusions in RQ2 on the ratios of unsafe functions and blocks in vulnerable packages.

\noindent\textbf{Conclusion validity.}
In RQ2, we investigated the affected versions of vulnerable packages. We noticed that some vulnerabilities do not explicitly indicate the earliest affected package versions, and considered the initial releases of their affected packages as the earliest versions affected by those vulnerabilities. The approach is also used in previous studies~\cite{zerouali2022impact} to estimate the earliest package versions affected by vulnerabilities, which could lead to inaccuracy in the range of affected package versions. To evaluate the inaccuracy of the approach adopted, we took a random sample of 40 vulnerabilities out of 364 that do not specify the earliest affected versions by vulnerabilities, with a 95\% confidence interval and 15\% sampling error. For each sampled vulnerability, we manually checked the corresponding fix commit(s) to determine whether the vulnerability exists in the initial release of the affected package. We found that 8 out of the sampled 40 vulnerabilities do not exist in the initial releases of the affected packages, leading to an overestimation of affected versions of packages by vulnerabilities. The 95\% Agresti-Coull confidence interval is (0.6498, 0.8976).

Memory safety issues account for the majority of disclosed vulnerabilities in the Rust ecosystem. The large amount of \emph{memory related} vulnerabilities may attribute to the fact that security experts and researchers tend to focus on memory problems in Rust packages given that Rust is claimed to ensure memory safety~\cite{Rudra,pldi2020,tosem2021}. Regarding vulnerability locality, security experts and researchers tend to focus on hunting vulnerabilities in unsafe code of Rust packages~\cite{Rudra,icse2020evan}, leading to the potential increase in the ratio of unsafe code in vulnerable code of the Rust ecosystem.

\section{Related work}\label{sec:related}
\subsection{Software Ecosystems}
Prior research presents a steady stream of empirical studies on various software ecosystems, including JavaScript (npm) ecosystem~\cite{msr2016npm,cogo2021empirical,kikas2017structure,zimmermann2019small,liu2022demystifying}, Python (PyPI)  ecosystem~\cite{hoving2013python,Bommarito2018PyPI,alfadel2021empirical,bagmar2021know}, and Java ecosystem~\cite{bavota2015apache,wang2020empirical, pashchenko2020vuln4real}.
In terms of the npm ecosystem, researchers investigate the package usage~\cite{msr2016npm, cogo2021empirical}, dependencies of packages~\cite{msr2016npm, zimmermann2019small,Cogo2021npm,kikas2017structure}, and security risks in the ecosystem~\cite{zimmermann2019small,liu2022demystifying}. The studies find that the npm ecosystem is steadily growing, with ongoing and accelerating growth in the number of packages and increasing dependencies between packages~\cite{msr2016npm}; individual vulnerable packages could impact a large portion of the entire npm ecosystem~\cite{zimmermann2019small}. 
As for the Python ecosystem, researchers characterize the developers of the ecosystem~\cite{hoving2013python}, growth in packages~\cite{Bommarito2018PyPI}, dependencies of packages~\cite{alfadel2021empirical}, and security risks in the ecosystem~\cite{alfadel2021empirical, bagmar2021know}. The findings indicate that the Python ecosystem grows exponentially~\cite{hoving2013python,Bommarito2018PyPI}; the number of disclosed vulnerabilities in Python packages increases over time, and  the vulnerabilities have been disclosed over 3 years after the relevant code was introduced in code repositories~\cite{alfadel2021empirical}. Studies on the Java ecosystem also explore the dependencies of packages~\cite{bavota2015apache}, and the security risks in the Java ecosystem~\cite{wang2020empirical, pashchenko2020vuln4real}.

A few studies investigate some aspects of the Rust ecosystem, by comparing it with other software ecosystems. Kikas et al.'s work~\cite{kikas2017structure} compared the structure and evolution of dependency networks across the JavaScript, Ruby, and Rust ecosystems. The reported results show that the analyzed ecosystems are alive and growing, with JavaScript having the fastest growth; software ecosystems are not as vulnerable to the removal of individual packages as they used to be.
Serebrenik et al.~\cite{serebrenik2015challenges} conducted a meta-analysis of the open challenges in software ecosystem research. As a result, they identified six open challenges, including quality and design, governance, dynamics and evolution, data analytics, domain-specific ecosystem solutions, and analysis of ecosystems.

Different from the studies above, we focus our research on the Rust ecosystem, by applying a systematic approach to investigating the security risks in the Rust ecosystem. 
Given dependencies of Rust packages have been investigated in a previous study~\cite{kikas2017structure}, we do not include such analysis in our study.

\subsection{Empirical Studies on Rust}
In the past, researchers have conducted empirical studies on various aspects of Rust libraries and projects in practice, including the usage of Rust language features ~\cite{astrauskas2020programmers, pldi2020} and bug characteristics~\cite{icse2020evan, tosem2021}. 
Some recent studies investigate the usage of unsafe Rust code in practice~\cite{astrauskas2020programmers,icse2020evan}. 
Astrauskas et al.~\cite{astrauskas2020programmers} analyze a large corpus of Rust projects to assess the validity of the Rust hypothesis,  and classify the purposes of unsafe Rust code. 
Evans et al.~\cite{icse2020evan} study how software developers use unsafe Rust code. 
Different from the studies on the usage of unsafe Rust code, our study investigates the relations between unsafe Rust code and disclosed vulnerabilities in the Rust ecosystem.
A recent study~\cite{schueller2022modeling} investigated the social risks of the Rust ecosystem from the perspective of developers and suggested ways for deploying limited developer resources to improve overall ecosystem health. Our work focuses more on the security risks in the Rust ecosystem from a technical perspective.

Other recent studies explore memory safety and concurrency issues in real-world Rust programs~\cite{pldi2020,tosem2021}. 
Qin et al.~\cite{pldi2020} manually inspect 850 unsafe code usages, 70 memory safety bugs, and 100 thread safety bugs located in five open-source Rust-based systems and applications, and five widely-used Rust libraries. 
Xu et al.~\cite{tosem2021} focus their study on 186 Rust bug reports related to memory safety by December 31, 2020.
They find that Rust can keep its memory safety promise, given that developers are unable to write memory-safety bugs without using unsafe code.

Different from Xu et al.'s work~\cite{tosem2021}, our study does not focus on memory safety issues. Instead, our study involves a variety of types of vulnerabilities and makes an in-depth analysis across different types of vulnerabilities. In addition, we investigate the relation between vulnerabilities and code regions with \textcolor{ao(english)}{\tt unsafe} keywords in vulnerable Rust packages.
Different from Qin et al.'s work~\cite{pldi2020}, our work investigates whether and how unsafe Rust code involves in vulnerabilities of the Rust ecosystem. Moreover, our dataset includes real-world Rust vulnerabilities of a variety of types, apart from the  memory safety and concurrency issues as investigated in the prior work~\cite{pldi2020}. 

\subsection{Securing Rust}
Some research efforts have been invested in securing Rust software via formal verification~\cite{reed2015patina, jung2017rustbelt, weiss2018rust, astrauskas2019leveraging, dang2019rustbelt, baranowski2018verifying}.
Patina~\cite{reed2015patina} is a formalization of the Rust type system. Patina captures the key Rust features relevant to memory safety and specifies how the features guarantee memory safety. 
RustBelt~\cite{jung2017rustbelt,dang2019rustbelt} defines rules to model Rust programs, and further uses these rules to prove the safety of Rust APIs. 
RustBelt also formally proves the memory safety of a realistic subset of Rust, including several standard Rust libraries with the existence of unsafe Rust. 
For a new Rust library that uses unsafe Rust code, RustBelt can tell the verification conditions that the library should meet to be considered as a safe extension to the Rust language.
Baranowski et al.~\cite{baranowski2018verifying} extend the SMACK verifier~\cite{rakamaric2014smack}, a software verification toolchain, to enable its usage on Rust programs. 
In addition, Astrauskas et al.~\cite{astrauskas2019leveraging} propose  a verification technique that utilizes the type system of Rust to simplify the specification and verification of Rust programs. The technique can assist developers to verify their programs with formal methods.

Researchers also propose numerous techniques to detect bugs in Rust software programs~\cite{toman2015crust,dewey2015fuzzing, lindner2018no}. 
Dewey et al.~\cite{dewey2015fuzzing} propose a fuzzing testing approach to detect type-checker bugs of Rust programs. Toman et al.~\cite{toman2015crust} introduce CRUST, a tool that combines exhaustive test generation and bounded model checking to verify memory safety in unsafe Rust code. The experimental results indicate that CRUST is effective at finding memory errors in the Rust standard libraries. 
Lindner et al.~\cite{lindner2018no} propose a verification process for Rust programs via symbolic execution to detect unsafe and panic issues. 
Our work investigates the fix patterns of different types of vulnerabilities in the Rust ecosystem to shed lights on securing Rust software programs.

\section{Conclusion and Future Work}\label{sec:conclusion}
In this paper, we conducted a large-scale empirical study on the security risks of the Rust ecosystem, one of the emerging and growing software ecosystems aimed at the development of systems software. 
Specifically, we characterized the disclosed vulnerabilities, vulnerable packages affected by the vulnerabilities, and corresponding  vulnerability fixes in the Rust ecosystem. 
We find that the vulnerabilities of different types differ widely in total numbers, disclosure and fixing duration, growth rates, and distributions across package categories in the Rust ecosystem. Among the 17 vulnerability types we identified, the memory safety and concurrency issues account for the majority of the disclosed vulnerabilities and demonstrate the fastest growth rates over time. One-third of the vulnerabilities have no fixes released by their public disclosure, leaving a window of opportunity for potential attacker exploitation. In the vulnerable packages, vulnerable code has statistically significantly higher ratios of unsafe functions and unsafe blocks compared to complete code, implying the potential higher security risks in unsafe functions and unsafe blocks. In addition, we identified three fix patterns for Rust vulnerabilities of different localities in Rust code.

Future work could consider developing an automatic tool to continuously collect and analyze the packages and vulnerabilities in the Rust ecosystem, and further leverage the real-time analysis results to gain awareness of the security risks in the Rust ecosystem in a timely way. 


\bibliographystyle{ACM-Reference-Format}
\bibliography{rust_ecosystem}


\end{document}